\documentclass{emulateapj}
\begin{document}
\title{The Relationship Between X-ray Luminosity and Major Flare Launching in GRS~1915+105}
\author{Brian Punsly\altaffilmark{1} and J\'{e}r$\hat{\mathrm{o}}$me Rodriguez\altaffilmark{2}}
\altaffiltext{1}{1415 Granvia Altamira, Palos Verdes Estates CA, USA
90274 and ICRANet, Piazza della Repubblica 10 Pescara 65100, Italy,
brian.punsly1@verizon.net or brian.punsly@comdev-usa.com}
\altaffiltext{2}{Laboratoire AIM, CEA/DSM-CNRS-Universit\'{e} Paris
Diderot, IRFU SAp, F-91191 Gif-sur-Yvette, France.}
\begin{abstract} We perform the most detailed analysis to date of the
X-ray state of the Galactic black hole candidate GRS~1915+105 just
prior to (0 to 4 hours) and during the brief (1 to 7 hour) ejection
of major (superluminal) radio flares. A very strong model
independent correlation is found between the 1.2 keV - 12 keV X-ray
flux 0 to 4 hours before flare ejections with the peak optically
thin 2.3 GHz emission of the flares. This suggests a direct physical
connection between the energy in the ejection and the luminosity of
the accretion flow preceding the ejection. In order to quantify this
concept, we develop techniques to estimate the intrinsic
(unabsorbed) X-ray luminosity, $L_{\mathrm{intrinsic}}$, from RXTE
ASM data and to implement known methods to estimate the time
averaged power required to launch the radio emitting plasmoids, $Q$
(sometimes called jet power). We find that the distribution of
intrinsic luminosity from 1.2 keV - 50 keV,
$L_{\mathrm{intrinsic}}(1.2 - 50)$, is systematically elevated just
before ejections compared to arbitrary times when there are no major
ejections. The estimated $Q$ is strongly correlated with
$L_{\mathrm{intrinsic}}(1.2 - 50)$ 0 to 4 hours before the ejection,
the increase in $L_{\mathrm{intrinsic}}(1.2 - 50)$ in the hours
preceding the ejection and the time averaged
$L_{\mathrm{intrinsic}}(1.2 - 50)$ during the flare rise.
Furthermore, the total time averaged power during the ejection ($Q$
+ the time average of $L_{\mathrm{intrinsic}}(1.2 - 50)$ during
ejection) is strongly correlated with $L_{\mathrm{intrinsic}}(1.2 -
50)$ just before launch with near equality if the distance to the
source is $ \approx 10.5$ kpc.
\end{abstract}

\section{Introduction}
Microquasars are Galactic X-ray binaries known to be sources of
relativistic jets \citep{mir92}. It is now clear that microquasars
exhibit at least two drastically different types of jets: discrete
ejections of 'plasmoids' \citet{mir94} usually associated with
transition from spectrally hard states to soft states (e.g.
\citet{fen04}), while a so-called compact jet is present during the
hard state (\citet{cor00,cor03,gal03}) and is quenched at the
transition to the soft state \citep{fen98}. If the accretion of
matter by the strong gravitational field due to the presence of the
black hole is natural, the ejection of particles at relativistic
speeds is mysterious, and still largely unknown. Huge
multi-wavelength efforts have been made in the past twenty years to
try and understand the origin of all type of jets and their
connection to the accretion processes (e.g.
\citet{mir98,cor00,cor03,kle02,gal03,rod08,rod09,rus11}). The
Galactic black hole candidate GRS~1915+105 is well known for
launching superluminal radio flares out to large distances
\citep{mir94,rod99,fen99,mil05}. In fact, there is no other known
Galactic black hole that has produced nearly as many strong radio
flares that have been observed to be superluminal. GRS~1915+105 is
known to show the two main types of jets \citep{mir94,dha00,fuc03}.
Discrete ejection, in this source, are further separated into to
superluminal ejections (resulting in major $\geq$ a few $100$~mJy
radio flares), and smaller 'bubble' ejections\footnote{The
discriminant between major flares and minor flares in our analysis
is based on our working definition of a major flare. Major flares
are defined as flares with radio spectra that evolve from being flat
spectrum at 15 GHz and lower frequency to being steep spectrum at
2.3 GHz with a flux density level of at least 100 mJy (where a two
point spectral index from 2.3 GHz to 8.3 GHz of 0.5 is used to
divide steep from flat spectrum). Typically, \textbf{if} a minor
flare is ever detected as steep spectrum at 2.3 GHz, the flux
density is less than 10 mJy.} showing sometimes a repetitive pattern
(radio oscillations; e.g. \citet{kle02,vad03}). The X-ray luminosity
of GRS 1915+105 is also one of the highest of any known Galactic
black hole candidate \citep{don04}. The existence of both
relativistic outflows and high continuum luminosity make it tempting
to speculate that GRS 1915+105 might be a scaled down version of a
radio loud quasar. This is of particular importance because the time
scales for evolution are reduced from the AGN (active galactic
nuclei) time scales of thousands to hundreds of thousands of years
to orders of seconds or hours. Thus, unlike quasars, it is in
principle possible to see a temporal connection between the putative
accretion flow (X-ray luminosity) and the superluminal jet launching
mechanism.
\par Studies have been performed in the case of the small
bubble ejections \citep{fen97,poo97,mir98,kle02,rod08}.
These ejections are always preceded/occur as response to a hard X-ray dip, with a
length that correlates with the radio flare amplitude, ended by a short soft X-ray spike
that is the trigger of the ejection \citep{rod08, rod09, pra10}.
\par More to the point, there are also anecdotal coarsely time sampled discussions of the
actual major flares. As an extrapolation of the discussion above on
X-ray cycles for optically thick flares, there is the analogy to the
major flares that has been perceived in some instances. The appeal
of these arguments is that it would unify all the radio behavior of
GRS~1915+105. X-ray dips in hard states preceding major flares that
occur on the order of days (as opposed to minutes as for the minor
flares) before the major flares start were noted in \citep{dha04}.
There is also mention of "associated" soft X-ray spikes that appear
when there are major flares \citep{tru07,fen04,dha04,vad03}.
The time relation between the soft spike and the radio
flare is, however, not as clear as for the bubble ejections.
Thus, considering the rapid rise of the
major flares (a few hours), we believe that a more contemporaneous
analysis is required in order to make a physical connection between
the X-ray emission and the major flare launching mechanism.
The only way to assess this is by means of a detailed study
of a large database of major flares.
\par Thusly motivated, this paper describes a program of study aimed
at determining the X-ray connection to major flares on the relevant
time scales of hours with existing long term archival data. Clearly,
minute by minute or second by second X-ray and wide-band radio
monitoring would tell us the answer, but it is not available and
impractical to collect. So we must settle for imperfect minute to hour
time resolution to get the best available insight.

\par The major flares in GRS~1915+105 occur at irregular intervals a few times a
year. The characteristic timescales for this system can be very
short, seconds to hours. Triggered observations only occur after the
bulk or all of the flare launching is over (eg, \citet{dha04}). Triggered radio observations of these flares
with MERLIN, VLA and VLBA and X-ray observations typically take place a day or
more after the ejection has occurred
\citep{mir94,rod99,fen99,dha00}. The major flares initiate unexpectedly on average
a few times a year. Thus, the actual initiation is very rarely
detected in the radio monitoring. Similarly, X-ray observations are only
serendipitous before unpredictable flare ejections and therefore are
generally too far spaced in time to be causally connected to the
launch mechanism. Thus, to our knowledge, all previous accounts of
X-ray connections to the major ejections have been based on time
sampling that is far too coarse ($\sim$ days) to uncover any causal
relationships between the state of the X-ray emitting plasma and the
powerful ejections.
\begin{table*}
\caption{Major Flares \tablenotemark{a}}
{\tiny\begin{tabular}{ccccccccc} \tableline\rule{0mm}{3mm}
Estimated Flare &  RXTE Date \tablenotemark{b} & $\Delta$ & End of Ejection & Estimated  & Observed & Type & $\alpha$ \tablenotemark{c} & Light\\
  Start (MJD) &  (MJD) & (hours) & (MJD) & Flux Density & Peak Optically & based& & Curve\\
  &  & & & 2.3 GHz & Thin 2.3 GHz & on & & Extrapolation\\
  &  & & & at $\tau=0.1$ & Flux Density & Figure 12 & & Time\tablenotemark{d}\\
  &  & & & (mJy) & (mJy) &&&(hrs)\\
\tableline \rule{0mm}{3mm}
50590.160 & 50590.157 (ASM)    & 0.05 & 50590.200  & $74 \pm 13$  & $ 63\pm 6$ & 1 & 0.57 & 0\\
$50750.520 \pm 0.02$   & 50750.351 (ASM)             & 4.53 & $50750.695 \pm 0.055$ & $490 \pm 58 $  & $ 548 \pm 55$ & 2 or 3 & 0.57 & 0.74-1.68  \\
$50916.140 \pm 0.04$  & 50916.089/50916.146 (ASM)\tablenotemark{e}    & 0.82 & 50916.49 & $702 \pm 129$  & $592 \pm 59$ & 1 & 0.52 & 4.80 - 6.72\\
50967.200              & 50967.148 (ASM)             & 1.26 & 50967.33 & $516 \pm 180$  & $343 \pm 34$ & 1 & 0.67 & 0 \\
51003.070             & 51002.907 (ASM)             & 3.91 & $51003.20 \pm 0.02$ & $234 \pm 29$  & $251 \pm 25$ & 2 &0.81 &0  \\
51270.330             & 51270.315 (ASM)             & 0.36 & $51270.45 \pm 0.04$ & $146 \pm 21$  & $145 \pm 15$  & 2 or 3 & 0.54 & 0.57 \\
51336.970             & 51336.940 (ASM)             & 0.72 & $51337.18 \pm 0.06$ & $470 \pm 49 $  & $484 \pm 48 $ & 2 or 3 & 0.70 & 0 \\
51375.150             & 51375.099 (ASM)             & 1.22 & 51375.28 & $196 \pm 25$  & $180 \pm 18$ & 1 & 0.60 & 0\\
51499.702             & 51499.614 (ASM)             & 2.11 & 51499.875 & $399 \pm 115$ & $358 \pm 36$ & 1 & 0.76 & 2.04\\
51535.630             & 51535.556 (ASM)             & 1.77& 51535.83 & $512 \pm 114$  &  $512 \pm 51$ & 1, 2 or 3 & 0.23\tablenotemark{f} & 0.93 \\
51602.506             & 51602.361 (PCA)             & 3.48 & 51602.63 & $125 \pm 13$  & $115 \pm 12$ & 1 or 3 & 0.80 & 0 \\
\end{tabular}}
\tablenotetext{a}{Catalogue of the major flares compliant to our
observational requirements (see text). } \tablenotetext{b}{The
closest (in time) RXTE X-ray observation date that preceded the
estimated flare ejection time. All data is with the ASM except for
the last entry which is with the PCA.} \tablenotetext{c}{The
spectral index when the peak optically thin 2.3 GHz flux density was
measured.}
\tablenotetext{d} {Length of time in hours of the gap in
coverage between radio data and the estimated start of the flare.}
\tablenotetext{e} {Depending on the exact flare initiation time, the
observation 50916.089 or 50916.146 could be the closest RXTE
observation that precedes the flare} \tablenotetext{f} {Technically
not optically thin, but see the caption of Figure 10 for an
explanation of the inclusion in the sample}

\end{table*}
\par Being cognizant of the discrepancy between the density of the time sampling
of data acquisition around the launch time of major flares and the major flare ejection time scale, a search for major flares
with high time resolution data sampling near the launch time was
initiated. We have found 11 major ejection events in which we have
sufficient information to assess the X-ray
state before and/or during flare launch and its connection to the
energy of the ejected plasmoids in major flares.
\par Our program has four main components, The first is finding
suitable major flare epochs that can be used for analysis as
described in Section 2. Secondly, we review and develop methods
for estimating the energy contained within a plasmoid from the time
evolution of the optically thin low radio spectrum (Section 3).
Thirdly, we determine conversion techniques in order to interpret
RXTE fluxes as intrinsic X-ray luminosity (Section 4). Finding an
algorithm to convert ASM fluxes to 1.2 - 50 keV intrinsic
luminosity, $L_{\mathrm{intrinsic}}(1.2 - 50)$, was the most labor
intensive effort and required attention to many technical points. In
the last portion of the program, we correlate the directly measured
quantities as well as the model dependent intrinsic physical
quantities in order to shed light on any underlying causal
relationships in the flare launching mechanism (Section 5).
\section{A Catalog of Major Flares with High Temporal
Resolution}Numerous studies of AGNs and GRS~1915+105 in particular
have indicated that optically thin low frequency radio emission is a
robust surrogate for determining the energy contained within the
plasma responsible for said emission \citep{raw91,wil99,pun12}. This
is discussed in detail in the next section where we make our
estimates of ejected plasmoid energy. In GRS~1915+105, it was
demonstrated that knowledge of the peak of the optically thin flux
at low frequency (2.3 GHz in this case) and the spectral index at
late times in the flare evolution removes many unknown variables
from the energy estimate when the data it is considered in the
context of an evolving plasmoid \citep{pun12}. The requirement of
sufficient data to implement this method greatly restricts the
available data set for our work. We find that only the GBI (Green
Bank Interferometer) survey has sufficient low frequency data time
sampling density at 2.3 GHz to define the peak of the optically thin
flux from an episodic ejection and show its late time behavior (The
GBI data was downloaded from the public access web sites:
ftp://ftp.gb.nrao.edu/pub/fghigo/gbidata/gdata/1915+105 and the
older data which was referenced but did not survive the culling
process that is defined below
http://www.gb.nrao.edu/fgdocs/gbi/arcgbi/1915+105). The GBI
monitoring ended in 2000. This motivated the following order of
culling criteria
\begin {enumerate}
\item A flare was loosely defined as 70 mJy of optically thin flux at
2.3 GHz, keeping in mind that the gaps in temporal coverage might
miss the true peak (which could very well be somewhat larger, $\sim$
100 mJy). The optically thin condition is defined by a spectral
index, $\alpha >0.5$ based on the convention that the flux density
follows a power law in frequency, $S_{\nu} \sim \nu^{-\alpha}$. An estimate of the peak
optically thin radio flux at 2.3 GHz allows us to estimate the
energy stored in the plamsoid. We use GBI monitoring data to minimize temporal gaps
in the coverage.
\item The next step was to use the 8.3 GHz flux density from the GBI survey and
15 GHz flux density from the Ryle Telescope monitoring program (see
the public archive http://www.mrao.cam.ac.uk/~guy/1915/ that has been generously provided by Guy Pooley) to find
epochs in which the initial period of optically thick emission that
precedes a corresponding optically thin peak at 2.3 GHz had
sufficient high frequency coverage to estimate the beginning of the
rise in optically thick flux.
\item We then applied a further constraint from the 8.3 GHz GBI data and the
15 GHz Ryle data that the peak of the optically thick
emission was either observed or fell within an observing gap less than $\sim$ an hour wide. Steps 2 and 3 provide a
rise time estimate and by 1), above, also an estimate of the time
averaged power required to eject the plasmoids.
\item The last step was the most critical. We looked through the
RXTE ASM and PCA (Proportional Counter Array) archives to find any
of the epochs that passed the first three steps of the culling
process that also had an X-ray observation to within 4 hours of the
initiation of a major ejection. In practice, RXTE pointing
observations that allow the exploitation of the PCA capabilities
(wide-band and large number statistics) rarely fall serendipitously
just before a major ejection. Thus, for the most part, the epochs
considered here have the more narrow range of coverage provided by
the three spectral band measurements of the ASM.
\end{enumerate}
\par The light curves in Figures 1 - 11, allow us to see the
estimates of the flare initiation time, the end of the flare
ejection, and the peak optically thin flux juxtaposed to the dates
of X-ray data capture. The horizontal and vertical scales on the
axes are optimized for estimates of flare rise times and locating
the peak optically thin flux density. In the light curves, one can
see that the slowly evolving peak in optically thin emission often
clearly lies within gaps in the GBI coverage. This leads to an
uncertainty in the peak optically thin flux. The value in column (5) is
the largest possible spread in the peak value that is
consistent with the observations. This is critical to our
calculations as the models of the December 1993 flares indicate that
this flux maximum occurs near the time that the plasmoid has an
optical depth $\tau \approx 0.1$. As such, we use this flux density
level (with its considerable uncertainty) as a surrogate for knowing the flux density when $\tau \approx
0.1$. The extremes of this spread in possible flux density when $\tau = 0.1$ are indicated by the black arrows in the light curves in
Figures 1 - 11. The error in column (5) represents the largest
deviation from the average of this spread. By contrast, column (6) is the largest optically thin 2.3 GHz flux
density taken directly from the measurements. The latter idea is
probably less accurate because it does not compensate for the large
gaps in observing coverage that are likely to miss the exact peak.
This cut the total number of major flares to just 11. Thus, only the
strongest correlations would be statistically significant from such
a small sample.
\begin{figure}
\begin{center}
\includegraphics[width=90 mm, angle= 0]{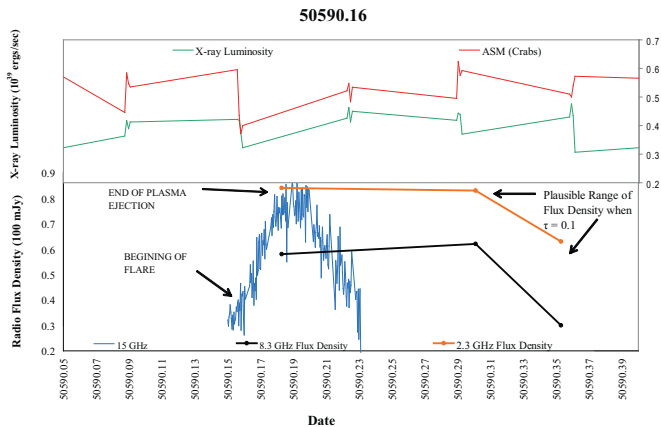}
\caption{\footnotesize{The top frame shows the X-ray light curves
for the flare that was initiated at approximately MJD 50590.16, the
ASM flux in Crabs and the intrinsic unabsorbed X-ray luminosity
estimated from the analysis in Section 4 (an assumed distance of 11
kpc is used throughout the paper unless otherwise stated) in units
of $10^{39}$ ergs/s. The radio data is plotted in the bottom frame
in units of 100 mJy. The 2.3 GHz coverage has gaps in it. A range of
plausible values for the flux density when the synchrotron self
absorption SSA optical depth = 0.1 are indicated by black arrows.
The method of designating this feature and the format of the split
frame figure are common to all the light curves presented in Figures
1 - 11. This is a weak flare by the standards of this paper and
might be a very strong version of a "bubble" flare or "baby jet."
Thus, it represents an important member of the sample because it
bridges the gap in power between the two classes of discrete
ejections that are discussed in the Introduction. Formally X-ray
observations precede the flare start time, but some X-ray
observations exist within a range spanned by the 10\% uncertainty
the was assigned to all flare start time estimates in Section 2.1.}}
\end{center}
\end{figure}
\begin{figure}
\begin{center}
\includegraphics[width=90 mm, angle= 0]{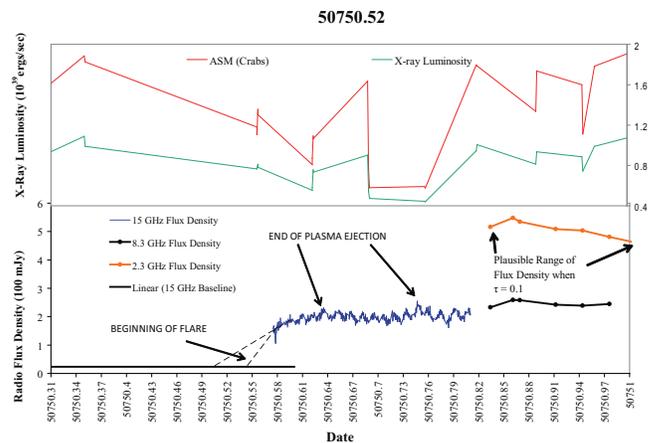}
\caption{\footnotesize{We perform two estimates of the flare
initiation time that are represented by the two dashed straight
black lines that extend from the 15 GHz light curve to the 15 GHz
background flux density level. There is significant curvature in the
light curve which leads to uncertainty in the extrapolation. The
steeper extrapolation is based on the linear least squares fit to
the first 10 minutes of 15 GHz data and the other extrapolation is
based on the fit to the first 30 minutes. This yields an estimated
initiation date MJD $50750.50 \pm 0.04$ for this flare. After the
first local maximum in the 15 GHz light curve at 50750.64, there is
a slight rate of increase, peaking again at 50750.74. Superimposed
on this are what appear to be strong core oscillations. Thus, the
true endpoint of the injection is obscured leading to significant
uncertainty. We carry the range of uncertainty in both the
initiation and end of the episode of plasma ejection throughout our
calculations.}}
\end{center}
\end{figure}
\begin{figure}
\begin{center}
\includegraphics[width=90 mm, angle= 0]{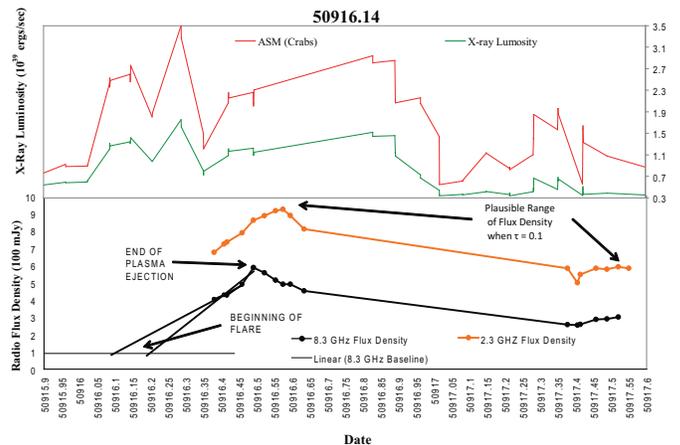}
 \caption{\footnotesize{The flare that was initiated at approximately MJD $50916.14 \pm 0.04$
is the most luminous at 2.3 GHz of any flare in the sample. The
extrapolation from the 8.3 GHz light curve to the flare initiation
time is the longest in the sample. Due to the sparse data and the
large extrapolation ($\sim$ hours), we consider two extrapolations,
one based on the first three points of the 8.3 GHz light curve and
the other based on all 4 points that comprise the rise in the flux
density. The extrapolations are represented by the straight black
lines. This results in uncertainty in the estimation of the flare
start time that manifests itself as a source of uncertainty in our
calculations of the power required to eject the plasma. The abrupt
change in the 8.3 GHz light curve from an increase to a decrease
seems to be a clear signal of the end of the energy injection as
opposed to adiabatic cooling.}}
\end{center}
\end{figure}
\begin{figure}
\begin{center}
\includegraphics[width=90 mm, angle= 0]{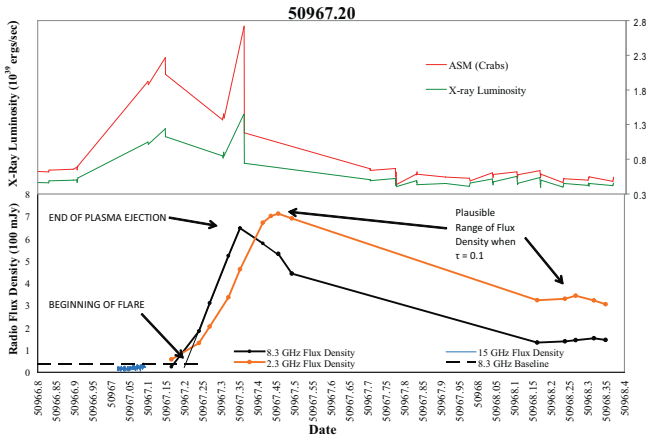}
\caption{\footnotesize{The flare that was initiated at approximately
on MJD 50967.20 is very strong with a very abrupt rise. There is a
large uncertainty in the peak optically thin flux density at 2.3 GHz
due to an inopportune gap in the GBI coverage. This creates a large
source of error in our energy estimates. The rise in the 8.3 GHz
light curve is very linear and its extrapolation (the straight black
line) appears to dive into a background level of random fluctuations
(crosses the dashed 8.3 GHz baseline flux density). The 8.3 GHz
light curve changes from a uniform linear increase to a decrease
very abruptly which seems to represent the ejection mechanism
turning off as opposed to gradual, continuous ageing or cooling
effects.}}
\end{center}
\end{figure}
\begin{figure}
\begin{center}
\includegraphics[width=90 mm, angle= 0]{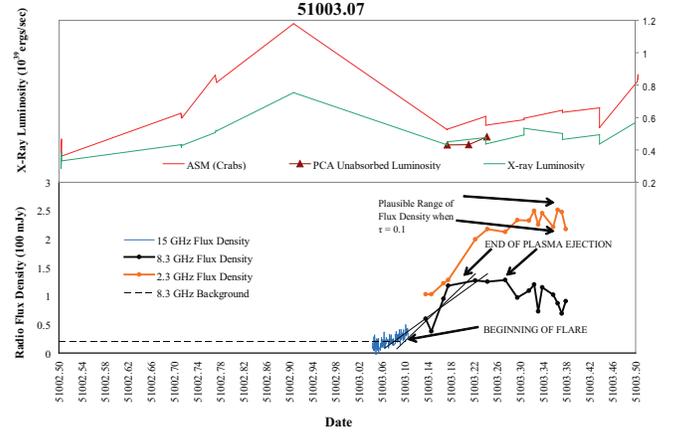}
\caption{\footnotesize{The flare that was initiated at approximately
MJD 51003.07 has both ASM and PCA X-ray data in the top frame. The
PCA data (discussed in section 4.4.2) was taken at what we estimate
to be the last few minutes to an hour of the plasma injection.
Notice how accurately the luminosity based on our narrow band ASM
estimator at this same time agrees with the broad band intrinsic
X-ray luminosity based on the detailed models of the PCA data. The
slow turnover following the abrupt rise makes the interpretation of
the end of the rise in the light curve ambiguous and introduces some
error. The rapid increase in the two point spectral index from 2.3
GHz to 8.3 GHz after the terminus of the steep rise of the 8.3 GHz
light curve is evidence of large optical depth changes due to
adiabatic expansion. The extrapolation of the 8.3 GHz light curve to
the background level (dashed black line) from 14 hours earlier has
some uncertainty due to large fluctuations during the rise. Two
representative fits (one a linear least squares fit to the first 5
points and one based on the first 6 points of the light curve) that
are consistent with the estimated flare end time are plotted as
black straight lines. The 15 GHz data has denser data sampling that
seems to straddle the flare initiation time and clearly provides a
better estimate of the start time.}}
\end{center}
\end{figure}
\begin{figure}
\begin{center}
\includegraphics[width=90 mm, angle= 0]{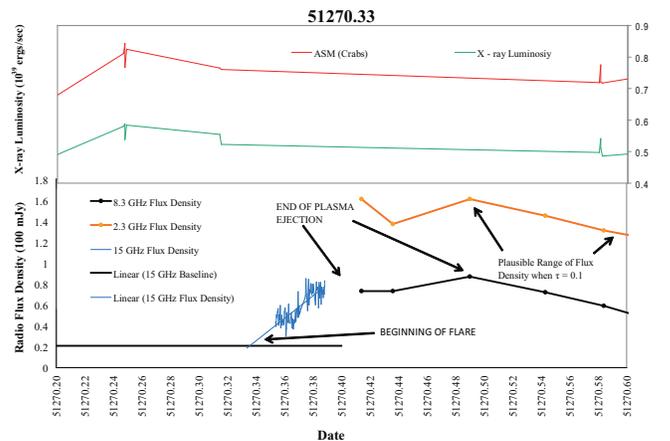}
\caption{\footnotesize{The flare that initiated on approximately MJD
51270.33 has a gap in the data coverage near the peak of the flare
rise at 15 GHz. The coverage resumes at 8.3 GHz which leaves a large
uncertainty in the estimate of the end of the plasma ejection
episode.}}
\end{center}
\end{figure}
\begin{figure}
\begin{center}
\includegraphics[width=90 mm, angle= 0]{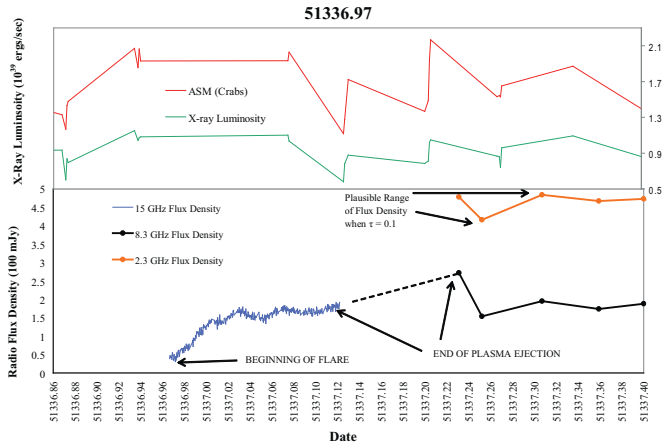}
\caption{\footnotesize{The flare that was initiated at approximately
MJD 51336.97 has excellent data sampling at 15 GHz during flare
initiation. The flare end time has some major uncertainty because of
strong oscillations superimposed on a slow rise in the light curve
from 51337.03 to 51337.13. Then there is a gap of 0.1 days until
coverage is resumed at 8.3 GHz which adds even more uncertainty. The
abrupt change in the 8.3 GHz light curve after 51337.231 seems to
indicate a distinct change in the plasma dynamics and the latest
possible estimate for the end of the flare. This flare and the VLBA
observations are discussed in detail in Section 2.2.}}
\end{center}
\end{figure}
\begin{figure}
\begin{center}
\includegraphics[width=90 mm, angle= 0]{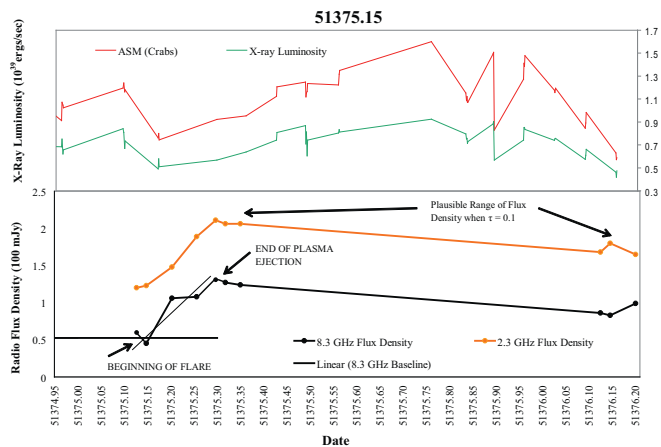}
\caption{\footnotesize{The flare that was initiated at approximately
MJD 51375.15 has complete 8.3 GHz coverage of the flare rise.
However, there is a large gap in the observation following this.
There is fortuitous ASM data (top frame) before and during the flare
rise. This is a modest flare that emerges from one of the highest
background flux levels of any flare. This combination of
circumstances manifests itself as a large uncertainty in the
spectral index of the optically thin emission in Table 2, columns 2
and 3.}}
\end{center}
\end{figure}
\begin{figure}
\begin{center}
\includegraphics[width=90 mm, angle= 0]{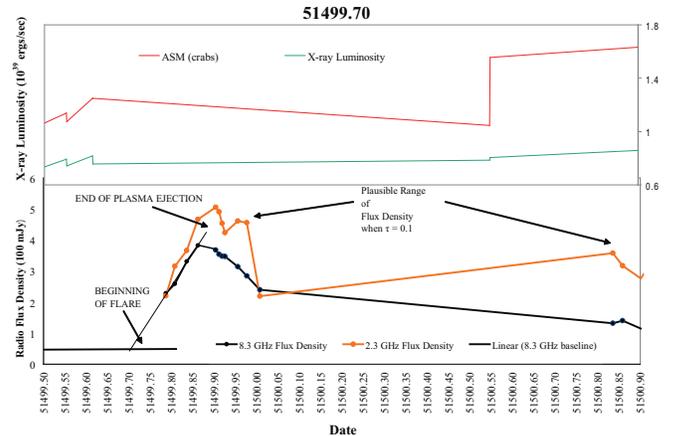}
\caption{\footnotesize{The strong flare that was initiated at
approximately MJD 51499.70 has good coverage all the way to the end
of the nearly linear steep rise of the 8.3 GHz light curve. The 2.3
GHz flux density drops abruptly near 51500.00. This type of drastic
dip is not seen in any other of the 2.3 GHz light curves. This "out
of family" behavior likely indicates a problem with the observation
or the data reduction.}}
\end{center}
\end{figure}
\begin{figure}
\begin{center}
\includegraphics[width=90 mm, angle= 0]{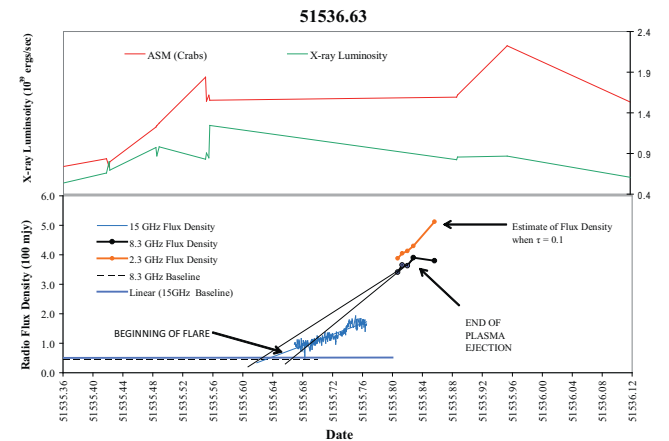}
\caption{\footnotesize{The strong flare initiated on MJD 51535.63
was included in the sample even though $\alpha <0.5$. Unfortunately,
the peak in the optically thin flux at 2.3 GHz was most likely
missed by a large gap in GBI coverage. The light curve had clearly
peaked at 8.3 GHz and was in the process of evolving towards being
optically thin between 2.3 GHz and 8.3 GHz when the data collection
stopped. We also have a good estimate of the flare start time from
the 15 GHz light curve and ASM data just before this. Since the
flare was very strong and the rise time could be estimated
accurately, we included it in the sample and just carried a large
error in the peak optically thin flux at 2.3 GHz. Due to the sparse
data at 8.3 GHz and the large extrapolation ($\sim$ hours) to the
flare start time, we consider two linear extrapolations, one based
on the first three points of the 8.3 GHz light curve and the other
based on all 4 points that comprise the rise in the flux density.
The extrapolations are represented by the straight black lines. The
15 GHz data has denser data sampling and requires a shorter
extrapolation to the initiation time and therefore provides a better
estimate of the start start time.}}
\end{center}
\end{figure}
\begin{figure}
\begin{center}
\includegraphics[width=90 mm, angle= 0]{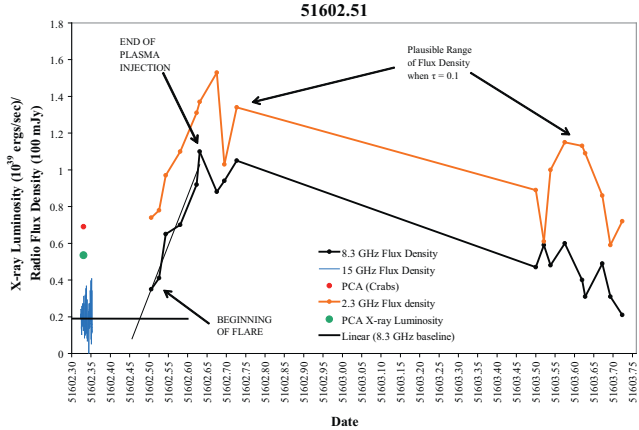}
\caption{\footnotesize{The light curves for the flare that was
initiated at approximately MJD 51602.506. This is the only epoch
with PCA observations immediately preceding the flare onset.}}
\end{center}
\end{figure}
\begin{figure}
\begin{center}
\includegraphics[width=90 mm, angle=0]{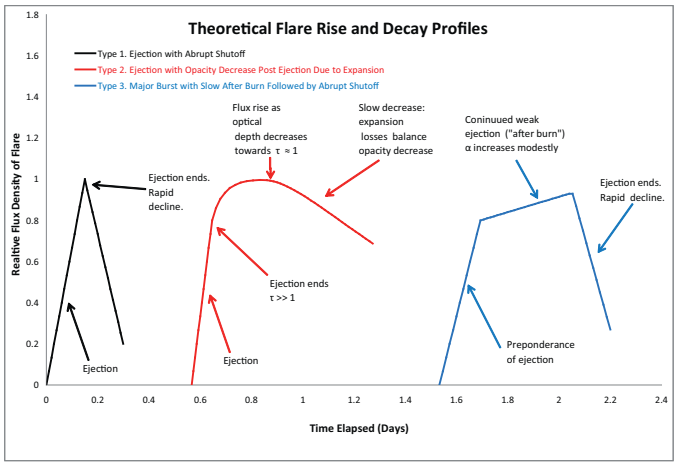}
\caption{\footnotesize{Generic light curves based on different
theoretical assumptions. See the text of Section 2.2 for details.}}
\end{center}
\end{figure}
\begin{figure}
\begin{center}
\includegraphics[width=70 mm, angle= -90]{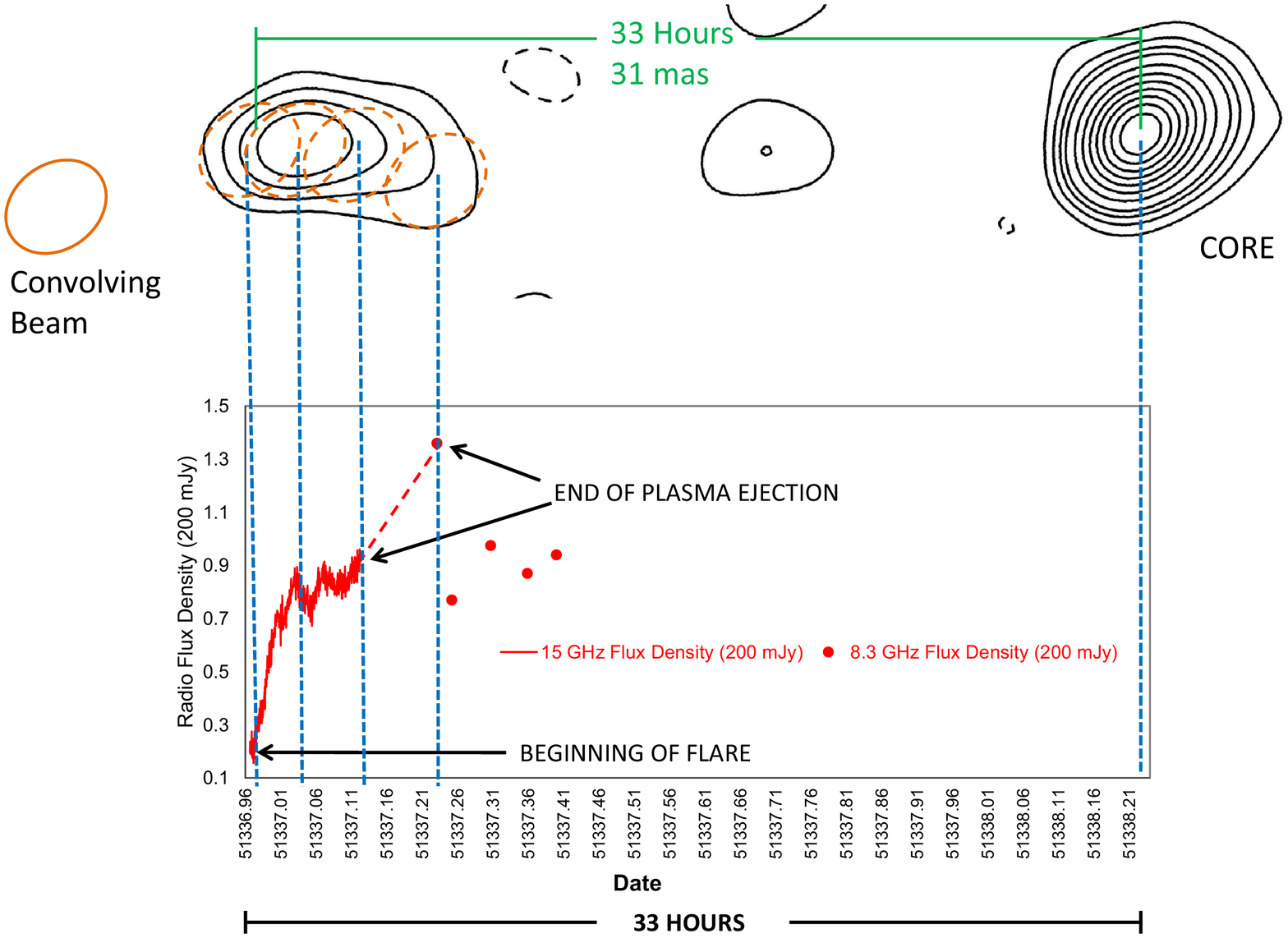}
\caption{\footnotesize{An 8.3 GHz VLBA image obtained 33 hours after
flare initiation. The red curve is the 15 GHz flux density and the
red dots are the 8.3 GHz flux density. The convolving beam used to
process the image from the VLBA observation is the red closed curve
with a size of 3.5 mas x 3.1 mas. The contour levels relative to the
peak flux of 32.2 mJy are at -5\%, 5\%, 10\%, 20\%, 30\%, 40\%,
50\%, 60\%. 70\%, 80\%, 90\%. The radio image was generously
provided by Vivek Dhawan. The emission between the core and ejection
in the image is an artifact of the time variable core combined with
sparse coverage in the u-v plane. See Section 2.2 for details.}}
\end{center}
\end{figure}

\subsection{Determining the Start of the Flare} Knowledge of the start of the flare is essential for
our analysis. It allows us to establish a temporal (and perhaps
causal) chain of events. It is this information that provides the
key time signature for the precise physical significance of
individual X-ray observations. Every optically thin flare is
preceded by a rise in optically thick high frequency radio emission.
As the ejected plasmoid expands, the optical depth to synchrotron
self absorption (SSA) decreases and the spectrum steepens at ever
decreasing frequency until it is optically thin at our frequency of
observation, 2.3 GHz. The highest frequency observations that we
have in general are the 15 GHz Ryle telescope data and 8.3 GHz GBI
data. We consider a rise in these data as an indication of a flare
initiation if it extrapolates in time to increased flux and
optically thin 2.3 GHz emission. \footnote{The optically thin nature
is critical since GRS 1915+105 can have extraordinarily strong
optically thick compact jets ($\sim$ hundreds of mJy), the so-called
"high plateau" state \citep{fuc03}.}
\par  In order to define the initiation of the
flare, we need to extrapolate the 15 GHz or 8.3 GHz light curve
backwards in time a few minutes to a few hours (i.e., there is not
complete temporal coverage). There are two considerations here that
are required for the sake of accuracy. First of all, we only select
flares where the light curve produces very short extrapolations in
time, typically less than one hour (see column 9 of Table 1).
Exceptions are only made for some of the extremely strong flares
that are desirable to study for their strength and for making
connections to previous published studies of these conspicuous
flares. Even so, the average extrapolation in Table 1 is less than
one hour. Secondly in order to execute this criteria, one must
define when the emission has actually begun to rise above the
background flux. We can establish the background flux by the level
of the flux in the previously measured epoch. So an additional
requirement for an epoch to have an acceptable data set is having
other measurements close enough in time (less than one day) before
the flare so that a background flux level can be estimated at 8.3
GHz or 15 GHz. When the extrapolated flux density level intersects
the background level, this is considered the start of the flare. The
background levels are usually between 10 mJy - 30 mJy for the 15 GHz
light curves and 20 - 40 mJy for the 8.3 GHz light curves. Because the
levels are so low and the rise is so steep this leads to minimal
errors in the estimations of the beginning of the flare rise. In
order to deal with this uncertainty, we liberally assign a 10\%
error to all our flare initiation time estimates and all our flare
end time estimates in the final calculation of ejection power.
Additional uncertainties in the flare initiation time due to the
details of the light curve data sampling are described in detail and
are quantified numerically and carried as an additional source of
error in all calculations.

\par It is important to justify the definition of the beginning of
the flare. Consider an optically thick source with an expanding
radius, $ r = f(t)r_{0}$, where $r_{0}$ is a constant and $f(t)$ is a scale factor
which for uniform expansion is proportional to elapsed time. If the
source is observed well below the SSA cutoff frequency,
\citet{mof75} showed that the spectral luminosity will scale in time
as
\begin{equation}
L_{\nu}(t) \sim f^{3}\;, \mathrm{for\, uniform\, expansion}\;
L_{\nu}(t) \sim t^{3}\;.
\end{equation}
Equation (1) is true for all frequencies well below the SSA cutoff frequency. Thus, the luminosity will
begin to increase at all frequencies that are well below cutoff,
simultaneously. However the rate of increase will be less for lower
frequencies due to the increased SSA opacity which scales like
$\nu^{-(2.5+\alpha)}$, where $\alpha$ is the spectral index
\citep{mof75}. Thus, all light curves that are based on optically
thick fluxes at a single frequency during the entire flare rise
should extrapolate back to the same flare initiation time. One
cannot combine fluxes at different frequencies in the same light
curve extrapolation. The higher the frequency, the quicker that the
flux density rises to levels above the background noise level making
the early stages of flare rise more discernible than at lower
frequency. This raises the question if using 8.3 GHz as opposed to
15 GHz will make a significant difference to the estimate of the
flare start. Using the light curves in Figures 4, 5, 10 and  11
allows us to assess this by estimating flare start times with both
8.3 GHz and 15 GHz (per the methods described in this section). For
the 8.3 GHz estimate we have a $-0.02$ day (lead) to $0.03$ day
(lag) offset for epoch MJD 51536.63, 0.00 to +0.03 days offset for
epoch MJD 51003.07, $<0.12$ days offset for MJD 50967.20 and $<0.15$
days offset for MJD 51602.51 (hereafter we will drop the MJD from
the dates for the sake of brevity). The upper limits represent the
gap in the data sampling, not a perceivable difference in flare
initiation estimates. This small uncertainty seems to be
accommodated by our choice of 10\% error on all flare initiation and
end time estimates (that was noted above).
\par Due to the short duration of the light curve extrapolations
in Table 1 and the frequency independent implications of Equation
(1), we conclude that light curve extrapolation is a robust
method of estimation that should be more accurate than extrapolating
ballistic motion from long baseline radio interferometry 53 to 77
hours after initiation as in \citet{fen99}. One reason for adding
the flare from epoch 50750.50 to the sample in spite of the
relatively long light curve extrapolation times of 0.74 - 1.68 hours
is the fact that this is a well studied flare, \citep{fen99,dha00},
and it is very strong. We can check the validity of our assumptions
and method by comparing the estimated initiation times. Using MERLIN
observations 53 to 77 hours later \citet{fen99} estimate an
initiation date of 50749.8 to 50750.08, we estimate 50750.00 to
50750.04. The agreement in the two methods and the fact that the
light curve extrapolation has less uncertainty is corroboration of
the discussion of the robustness of this method that was given above. We
expect estimates with shorter extrapolations to be equally or more
accurate.

\subsection{Determining the End of the Plasma Ejection Episode} Knowing both the
beginning and the end of the ejection episode is a valuable piece of
information as it allows for a conversion of the energy in the
ejection to the time averaged power required to eject the plasmoid.
Since the existence of an increase of optically thick high frequency
emission (before an optically thin flare) is the signature of
plasmoid ejection, we consider the most straightforward
interpretation of the light curves: the maximum of the optically
thick emission at 8.3 GHz or 15 GHz signals the end of the event
corresponding to the preponderance of plasma ejection. This is an
important assumption in the following and we must look at it
critically and within the context of the data available for these
flares in particular. In general, one must note the following
fundamental ambiguity. The attenuation due to optically depth
evolves far more rapidly than adiabatic cooling as was shown in
\citet{pun12}. So, an optically thick plasma in a state of expansion
that is becoming optically thin at a specific frequency may tend to
increase in spectral luminosity even though there is no injection of
energy \citep{mof75}. Furthermore, since the synchrotron self- absorption (SSA) optical
depth is frequency dependent this can lead to a frequency dependent ambiguity to the end
of the ejection. Yet since this process is distinct from the ejection process, it
would typically be accompanied by a significant change in the rate
of increase, curvature or slope of the radio light curve adjacent to
its maximum. In order to explore the nature of the rise and the
end of the ejection, we introduce three distinct types of light
curves below as a means of exploring the issues raised above.
\par These concepts can be
visualized in terms of the theoretical light curves that are
presented in Figure 12. The type 1 light curve (on the left hand side
of Figure 12) is the most common type in the sample described by
Table 1. The discussion of equations (2) - (6) below indicate that there is only one reasonable interpretation of this
light curve - some physical process (probably the ejection
mechanism) ends abruptly. There is very little ambiguity in this
instance. It should be noted that the SSA opacity is not typically that large in such a
sceanrio when the light curve turns over. So changes in the attenuation occur at
a much slower rate than changes in the radiative efficiency. This is evidenced by most
of the light curves which often have relatively steep spectral indices ($\sim 0.5$)
between 2.3 GHz and 8.3 GHz when the high frequency light curve turns over (eg., Figures 3, 5, 6, 8, 9 and 11).
\par The decrease in the SSA optical depth resulting from plasmoid expansion
can in principle be the dominant contributor to the luminosity
increase near the maximum. This scenario is depicted in the middle
light curve of Figure 12, the type 2 flare. The gradients in the
light curve curvature near the maximum are distinct from a type 1
flare. A decrease in optical depth causes the luminosity of an
expanding plasmoid to change gradually (semi-continuously). This has
two manifestations not seen in the type 1 flare. The first is a
flattening of the rise in the spectral luminosity light curve.
The second is a gradual roll-over in the light curve at later times that
transitions to a spectral luminosity decrease. These general
characteristics of the curvature are independent of the detailed
model of the rate of expansion and distribution of plasma.
 \par The type 3 light curve is subtly different from the type 2 light curve (Figure 12). The physical mechanism
  responsible for this light curve is very different though. The first inflection point is the zeroth order realization
  of a drastic change in the ejection rate. This can be described as a powerful major ejection followed by a low level
  much weaker ejection that can last many hours, a "slow after burn". After which, the ejection shuts off and
  there is a rapid decrease in spectral luminosity as for the type 1 flare. Another difference between the
  type 2 and type 3 flares is that type 2 flares (based on the hypothesis of strong optical depth changes as the root cause of the
  spectral luminosity increase at 8.3 GHz) should be accompanied
  by very rapid and continual spectral index changes between 8.3 GHz and 2.3 GHz. By contrast, the spectral index changes
  are much less during the late time portion of the rise in the light curve for type 3 flares, since there is continual low level
  injection of optically thick plasma.
\par What is particularly striking about most of the light curves in Figure
1 - 11 is the extremely linear rise in flux density. Excellent
examples are the epochs 50590.15, 50967.20, 51003.07, 51270.33,
51336.97, 51536.63, 51602.51 - 7 out of the 11 epochs. Note that
this linear increase combined with equation (1) is in conflict with
a scenario of nearly instantaneous injection (i.e time scales for
injection much shorter than the rise time). For an instantaneous
ejection and free expansion, one would expect a much more rapid
($\sim t^{3}$) increase in the flux density after the flare
initiates. It is also curious that 5 of the epochs show clear signs
of a discontinuous change (an abrupt change on time scales less than
one hour) in the light curve from a linear rise to an approximately
linear decrease (see Figures 1, 3, 4, 8 and 9). This abrupt, almost
discontinuous behavior that typifies $\sim$ half of the flares is
also not consistent with the adiabatic expansion of an instantaneous
injection. One expects a gradual transition form rise to decay in
the expanding instantaneous plasmoid injection scenario in general
\citep{mof75}. It is possible for the light curve decline to appear
rather abrupt if the plasmoid is "very small" and the peak frequency
is "very high" in a pure adiabatic expansion (of an instantaneous
injection) scenario as noted in Figure 1 of \citet{van66}. However,
there is a necessary consequence to assuming such small sizes and a
"high" frequency light curve maximum to explain the abrupt changes
seen in the 8.3 GHz light curves. Namely, the light curve maxima at
lower frequency (that occur at later times) are of significantly
lower flux density than the narrower, earlier, high frequency light
curve maxima as noted in \citet{van66,pun12}. But, this is not the
case for the major flares in GRS 1915+105 as evidenced by the light
curves in Figures 3, 4, 5, 8, 9, 10, and 11 in which the opposite
occurs. The time delayed low frequency (2.3 GHz) light curve maxima
have larger flux densities than the earlier, higher frequency light
curve maxima (8.3 GHz). Thus, pure adiabatic expansion scenarios are
at odds with the spectral evolution of the major flares. The
observed spectral evolution is consistent with the notion that 8.3
GHz is too low of a frequency to be associated the peak frequency of
the SSA spectrum in early stages of major flare evolution in GRS
1915+105 (when the plasmoid is "very small") as discussed in
\citet{pun12}. The fact that the light curve maxima increase in
amplitude at lower frequency (contrary to adiabatic expansion) was
shown to be a result of the radiative efficiency increasing, as the
minimum energy condition is approached, during the December 1993
flares \citep{pun12}.
\par There is a simple explanation
of this abrupt inflection in terms of a source that turns off. To see this,
define the spectral luminosity in terms of the total volumetric
efficiency (which includes the effects of local emissivity and SSA
optical depth), $\epsilon_{total}(t)$ and the total internal energy
of the plasma (magnetic and leptonic), $E_{total}(t)$,
\begin{equation}
L_{\nu}(t) \equiv \epsilon_{total}(t)E_{total}(t) \;.
\end{equation}
For most flares it is verified that there is an approximately linear increase after
the start time, $t_{0}$, which we model in a simple mathematical form

\begin{equation}
L_{\nu}(t) = A (t-t_{0}) \;,
\end{equation}
where $A$ is a constant. The total energy as a function of time can
be described by a source, $Q(t)$ (injected power) and a sink $S(t)$
(i.e radiation losses, expansion losses).
\begin{equation}
E_{total}(t) = \int_{t_{0}}^{t} Q(T)-S(T)\, dT \;.
\end{equation}
Combining Equations (2) - (4) one finds
\begin{equation}
L_{\nu}(t)= A (t-t_{0}) = \epsilon_{total}(t)\int_{t_{0}}^{t}
Q(T)-S(T)\, dT \;, \; t < t_{o}
\end{equation}
For the numerous type I flares in Table 1, there is an abrupt change
in the functional form of the light curve at $t=t_{1}$ (as discussed
earlier in this section) which we model in a simple mathematical
form
\begin{widetext}
\begin{eqnarray}
L_{\nu}(t)= A (t_{1}-t_{0}) - B (t-t_{1})=
\epsilon_{total}(t)\left[\int_{t_{0}}^{t_{1}} Q(T)-S(T)\, dT
-\int_{t_{1}}^{t} Q(T)-S(T)\, dT \right]\;,
\end{eqnarray}
\end{widetext}
where B is a constant. There is no straightforward physical reason
for $\epsilon_{total}(t)$ or $S(t)$ to change abruptly (eg., in the case of instantaneous injection) since
synchrotron cooling, expansion losses and SSA opacity changes due to
adiabatic expansion are gradual \citep{mof75}. Thus, the
straightforward solution is that $Q(t> t_{1}) \ll Q(t< t_{1})$, ie.
the source has virtually shutoff. This does not eliminate more
elaborate scenarios as viable, but since there is a simple direct
interpretation of the data, we consider it reasonable, and look for
more corroboration of this interpretation below in the VLBA data. As
another note on this calculation, for type III flares this argument
applies to the second inflection point in the light curves. As for
the first inflection point, it suggests that $Q(t)$ decreases
significantly, but not to almost zero.
\par The abrupt cutoff from $Q(t)$ shutting off will produce a change
in the light curve independent of frequency by Equation (6).
Although the shape of the light curve around this inflection point
might be affected by the SSA opacity. For the type II flares the
profile is strongly affected by the frequency dependent SSA opacity
making the exact time of the source turnoff difficult to extricate
from the light curve. Hence, all type II flares carry an uncertainty
in the flare end date in Table 1 and the type I flares do not.
\footnote{The flare model in Section 3 and in \citet{pun12} is not
the shock model of \citet{mar85} which can create sharp inflection
points in the light curve by adjusting the numerous free parameters
\citep{tur00}. If the background jet is destroyed during the
ejection as indicated in \citet{dha04}, there is no jet for the
shock to propagate along.}
\par It is worth noting an additional complication that can potentially make the light curve interpretation
  as one of the three types a little bit ambiguous. After the major ejection, there can be strong optically thick
  shocks in the plasmoid \citep{dha04}. Furthermore, it was found in \citet{pun12} that there are often multiple ejections of compact components that
  radiate spectra that are peaked at high frequency. They are sufficiently weak that by the time they become optically
  thin at low frequency, they are too weak to be segregated from other weak sources in the system. If there are strong oscillations
near the end of the flare rise (eg., Figures 2 and 7) then
this creates significant uncertainty in the estimate of the end of
the plasma injection for that major flare.
\par We consider an anecdotal example closely in order to check
the details of our light curve surrogate for determining plasma
ejection durations. Fortunately, there is excellent data for the
flare that initiates at epoch 51336.970. Figure 13 shows a VLBA
8.3 GHz image taken 33 hours after the start of the flare that is
superimposed on the radio light curve. The center of the radio core
is placed at 33 hours after the flare initiation on the horizontal axis of the
plot. The "radio jet" is directed towards the flare start to show
its formation in a time reversed sense. In order to see if the
regions of peak flux density correspond to the rise of the optically
thick flux, we assume that the plasmoid propagation speed is
constant since inception. This was shown to be the consistent with
the time snapshots in other ejections imaged with VLBA and MERLIN
\citep{dha00}. So it is a reasonable assumption to implement here.
Thus, the length of the jet on the sky plane can be used as a ruler
in time, a linear transformation of the time axis, $T$, to the
angular separation from the core in the sky plane, $AS$. The simple
formula, for the time of ejection from the core is $T=AS/ V$
where $V$ = (31 mas/ 33 hours) = 0.94 mas /hour (which is just the
propagation speed of the jet on the sky plane). The closed contours
superimposed on the image in Figure 13 represent the convolved
beam-width that is used to create the image and can be considered
the resolution limit of the image. Amazingly, the strong peaks in
the flux density correspond with the rise in the 15 GHz flux density
in the light curve. The weak tail (jet) corresponds to the slight
rise in the 8.3 GHz flux density seen after this in the light curve.
It is unclear if this slight rise in the optically thick 8.3 GHz
flux and the weak tail is more ejected plasma or just back flow (in
the frame of the plasmoid) from MHD (magnetohydrodynamic) induced viscosity at the outer
boundary of the advancing plasmoid (eg., Kelvin-Helmholtz and
Rayleigh-Taylor instabilities) or back reflecting shocks from the
plasmoid interface with the interstellar medium as expected from the
method of characteristics \citep{san83}. This perfect alignment of
the light curve and the VLBA image may or may not be coincidental.
\par Even taking a pessimistic view that the alignment
in the figure is coincidental, one can still ask what the figure
indicates. Because of the extrapolation from the end of the 15 GHz
rise to the 8.3 GHz point in the light curve is very steep, one does
not know if the rate of increase of optically thick flux flattened
out around 51337.100 or not. The abrupt change in the 8.3 GHz light
curve after 51337.231 seems to indicate a distinct change in the
plasma dynamics. This is the maximum viable time for the end of
significant plasma ejection based on an analysis of the light curve
only and seems to be indicated by the VLBA image also. Therefore,
the radio image seems to independently corroborate the idea that
there is no major plasma ejection after the light curve has peaked
or has flattened out - we use the radio image to corroborate the
maximum limit on the the length of time plasma ejection that is
listed in Table 1.
\par The radio image can be used to shed some light on the
estimation of the lower range of the duration for the plasma to be
ejected. The 15 GHz light curve rises at a constant steep rate until
it reaches a local maximum at 51337.031. This aligns perfectly with
the peak of the VLBA image. However, the rise is also steep from
51337.100 and 51337.120. Thus, there seems to be significant plasma
ejection during this time interval as well. There is no clear
evidence that the 15 GHz light curve is flattening out at this point
as it would be doing if it were dominated by expansion effects
(i.e., there is insufficient data to distinguish between a type 2 or
type 3 light curve). The interpretation of an increase in ejected
plasma is consistent with the VLBA image, since the detected flux in
the image at the corresponding time is based on contours that are 3
and 4 times the noise level (as indicated by the magnitude of the
negative contour). The radio image supports the notion that a
significant if not the majority of the plasma was injected before
51337.031. But, it would be an enormous extrapolation of the
existing data to assume that the entire flux was ejected during this
time interval and all the extended structure in the VLBA is entirely
a consequence of the MHD effects spreading out the plasma
distribution as discussed above.
\par At a minimum, the
figure bounds the flare rise time between the end of the rise of the
15 GHz light curve and the peak of the 8.3 GHz for this flare. We do
not expect the interpretation of the light curve shape to be
significantly different for the other flares. In general, we carry
this uncertainty in the end of the flare rise time in all of our
calculations.
\section{Implications of the Optically Thin Low Frequency Radio
Emission}The implementation of low frequency radio flux ($\sim$ 151
MHz - 408 MHz) to estimate the energy content of the plasma within
radio lobes has a rich tradition in the study of radio loud AGN
\citep{raw91,wil99,blu00,pun05}. The low frequency is chosen to
eliminate significant contributions due to the optically thick
emission (synchrotron-self absorbed emission) from the radio core.
The fundamental uncertainties in the application of these methods to GRS~1915+105 are, \citep{fen99,pun12},
\begin{enumerate}
\item Is the plasma protonic or positronic?
\item What is the minimum electron energy, $E_{min}$ ($E_{min}=1$ in units of the electron rest mass energy, $m_{e}c^{2}$)?
\item There is uncertainty in the size of the region that produces the bulk of the radio emission.
The size of the physical region that produces the preponderance of the radio
emission in major flares appears to be smaller than the FWHM (full
width at half maximum) beamwidth of the VLA, VLBA or MERLIN
interferometers \citep{rod99,dha00,fen99}.
\item Is the minimum energy or the equipartition assumption justified?
\end{enumerate}
\subsection{Method of Estimating the Power Required to Eject Major
Flares} There are significant uncertainties listed above, many of which
are ameliorated in a method developed in \citet{rey09}. The modeling
method exploits the fact that synchrotron self-absorbed plasmoids
are restricted in physical dimension by the shape of their spectrum.
In particular, the frequency and the width of the spectral peak.
This provides two added pieces of information beyond the spectral
index and flux density of the optically thin high frequency tail.
Such an analysis provides strong constraints on the size of the
emitting region. In order to implement the modeling method, one must
have sufficient frequency coverage so that the peak and the high
frequency tail of the spectrum are defined. Furthermore, the data
must be quasi-simultaneous because strong flares evolve rapidly,
especially in their compact self-absorbed phase. A literature search
revealed one such instance of broadband simultaneous frequency
coverage in \citet{rod95} for a flare in December 1993. The modeling
of the self-absorbed components of this powerful flare and their
time evolution was studied in detail in \citet{pun12}. The insight
provided by the fortuitous monitoring of this very compact and
powerful flare motivates the assumption that the low frequency
optically thin radio emission should be a robust estimator of the
ejected plasma energy for other flares if it is used in conjunction
with knowledge of when the flare evolved from optically thick to
optically thin.

\par The relevant aspect of GRS~1915+105 discussed in \citet{pun12} was the time evolution of
the compact flares which greatly enhances the accuracy of already
insightful methods of plasmoid energy estimates developed in
\citet{rey09} and resolves most of the ambiguity noted at the beginning of this section. The
fact that the flares were very compact and strong allowed for a very
slow evolution of the synchrotron self absorption (on the order of a
few days). This facilitated adequate time
resolution to deduce the following behavior. The flares emerged
optically thick. The 15 GHz - 22 GHz emission rose because of the
injected energy and became enhanced as the ejected plasmoid became
less optically thick to synchrotron self absorption as it expanded.
Eventually the plasma became optically thin, even between 1.4 GHz
and 2.3 GHz. Physically, the plasmoid began very far from
equipartition and was magnetically dominated and positronic in
composition. As it expanded, it evolved toward equipartition and the
minimum electron, $E_{min} \approx 1$. Around the time that the
optically thin flux at 2.3 GHz was near a maximum, the plasmoid was
close to equipartition.
\par Our basic assumption is the following: the
detailed modeling of the time evolution of the flares from December
1993 can be used as a template for the time evolution of other
plasmoids that evolve far more rapidly and have sparser low
frequency radio coverage. The energy in the ejected plasmoids in
this paper are estimated by a method that is based on the optically
thin 2.3 GHz flux density. It is important to emphasize the
optically thin character for two reasons. First, for strong flares
the 2.3 GHz flux sometimes peaks when the flare is still optically
thick at 2.3 GHz and secondly the estimate on plasmoid size will
only be valid near the peak optically thin flux at low frequency. In
particular, we know the time evolution is one which begins with a
synchrotron self-absorbed optical depth of $\tau(2.3 \,\mathrm{GHz})
\gg 1$, corresponding to an attenuation factor of $e^{-\tau}\ll 1$.
As the plasmoid expands, the $\tau(2.3 \,\mathrm{GHz})$ decreases
and is less than unity when it is optically thin at this frequency.
Based on the models in \citet{pun12}, one expects $\tau(2.3
\,\mathrm{GHz}) \approx 0.1$ when the flux reaches its peak value.
\par The calculation proceeds as in \citet{pun12} and \citet{rey09} with
two equations that must be solved simultaneously. The first is an
equation for the optical depth,
\begin{equation}
\tau(2.3 \,\mathrm{GHz}) \equiv \mu(\nu =2.3 \mathrm{GHz})R=0.1 \;,
\end{equation}
where $\mu(\nu)$ is the synchrotron self-absorption (SSA)
attenuation coefficient (see equation (14), below) and R is the
radius of the plasmoid. The second equation is the flux density at 2.3 GHz when $\tau
=0.1$, $S_{\nu=2.3\mathrm{GHz}}(\tau =0.1)$, that can be approximated
from our knowledge of plasmoid evolution in \citet{pun12} as
\begin{widetext}
\begin{equation}
S_{\nu=2.3\mathrm{GHz}}(\tau =0.1)\approx \mathrm{peak \;
optically\; thin \; flux \; density \;at \;2.3\; GHz}\;.
\end{equation}
\end{widetext}
Equations (7) and (8) are solved simultaneously in a time snapshot
when the evolving plasmoid composition of the major flares in
\citet{pun12} is characterized by the constraints
\begin{enumerate}
\item the plasma is primarily positronic
\item the plasma is near a minimum in energy
\item $E_{min}\approx 1$.
\end{enumerate}
The solution is unique.
\par To convert these equations to the physical
parameters in the plasmoid, we assume a power-law energy
distribution for the relativistic electrons,
\begin{equation}
N(E)= N_{\Gamma}E^{-n} \;,
\end{equation}
where the radio spectral index $\alpha = (n-1)/2$ and $E$ is the
energy of the electrons in units of $m_{e}c^{2}$.
Consider the leptonic thermal energy density, $U_{e}$, and the
magnetic field energy density, $U_{B}$. Equation 9 with condition 3
above yields
\begin{equation}
U_{e} \approx m_{e}c^{2} \frac{N_{\Gamma}}{n -2} \; ,
\end{equation}
and
\begin{equation}
U_{B} = \frac{B^{2}}{8\pi} \; ,
\end{equation}
The total energy density stored in the leptonic plasma is given by
\begin{equation}
 U = U_{B} + U_{e}\;.
\end{equation}
If $V$ is the volume of the plasmoid, in the approximately uniform limit, we
define the total plasmoid energy as
\begin{equation}
E_{\mathrm{total}} \approx U V \;.
\end{equation}

\par To implement equation (7), we take the
standard result for the SSA attenuation coefficient in the plasma
rest frame (noting that $\nu = \nu_{o} / \delta$) from
\citet{rey96,gin69},
\begin{widetext}
\begin{equation}
\mu(\nu_{o})=\frac{3^{\alpha +
1}\pi^{0.5}g(n)e^{2}N_{\Gamma}}{8m_{e}c}\left(\frac{eB}{m_{e}c}\right)^{(1.5
+
\alpha)}\nu_{o}^{-(2.5 + \alpha)} \delta^{(2.5 + \alpha)}\;,\\
\end{equation}
\end{widetext}
\begin{equation}
g(n)= \frac{\Gamma[(3n + 22)/12]\Gamma[(3n + 2)/12]\Gamma[(n +
6)/4]}{\Gamma[(n + 8)/4]}\;.
\end{equation}
The Doppler factor, $\delta$, is given in terms of
$\Gamma_{\mathrm{rel}}$ (not to be confused with the gamma function
in the expression above), the Lorentz factor of the outflow;
$\beta$, the three velocity of the outflow and the angle of
propagation to the line of sight, $\theta$;
$\delta=1/[\Gamma_{\mathrm{rel}}(1-\beta\cos{\theta})]$
\citep{lin85}. In order to describe the total energy in
equation (13) in terms of observable quantities, we need to express
the spectral luminosity in terms of observed flux density. One can
express the observed flux density, $S(\nu_{o})$, in the optically
thin region of the spectrum using the relativistic transformation
relations from \citet{lin85},
\begin{eqnarray}
 && S(\nu_{o}) = \frac{\delta^{(k + \alpha)}}{4\pi D_{L}^{2}}\int{j_{\nu}^{'} d V{'}}\;,
\end{eqnarray}
where $D_{L}$ is the luminosity distance and $j_{\nu}^{'}$ is
emissivity evaluated in the plasma rest frame at the observed
frequency. The constant $k$ is of geometrical origin and is 3 for
unresolved emission as is the case here. To make the connection
between the observed flux density in equations (8) and (16) to the
local synchrotron emissivity within the plasma note that the
synchrotron emissivity is given in \citet{tuc75} as
\begin{eqnarray}
&& j_{\nu} = 1.7 \times 10^{-21} (4 \pi N_{\Gamma})a(n)B^{(1
+\alpha)}(4
\times 10^{6}/ \nu)^{\alpha}\;,\\
&& a(n)=\frac{\left(2^{\frac{n-1}{2}}\sqrt{3}\right)
\Gamma\left(\frac{3n-1}{12}\right)\Gamma\left(\frac{3n+19}{12}\right)
\Gamma\left(\frac{n+5}{4}\right)}
       {8\sqrt\pi(n+1)\Gamma\left(\frac{n+7}{4}\right)} \;.
\end{eqnarray}
Our solution is a numerical iterative method that solves equations
(7) - (18) simultaneously under the assumption of a spherical
volume. Without any knowledge to the contrary, we assume spherical
and homogeneous since the plasmoids are too small to be resolved \citep{pun12}. Per
condition "2" above, the parameters (N, B, R) are optimized until a solution to equations (7) to (18) with the
minimum energy is found. This is considered the physical solution and its
parameters are displayed in Table 2.
\subsection{Estimating the Number Density Spectral Index}
For a given kinematically determined $\delta$, the one unknown in the equations above is the number density spectral index,
$n$. This is determined from $\alpha$ of the optically thin component. First, one must
subtract the flux density of the optically thick core. We cannot
use the baseline flux density in Figures 1 - 11 to estimate this quantity
because the core jet gets disrupted during the ejection and
re-establishes itself at a different level after the ejection stops \citep{dha00,dha04}.
Since there is uncertainty associated with this quantity, we estimate the spectral
index in the presence of background core flux by two different methods
that we describe below.
\par The first method assumes that the random variations in the 8.3 GHz light curve that occur after the flare has
risen to its peak represent optically thick time variations of the core. This
estimation is computed by first performing a linear fit to the 8.3 GHz flux density over a 3 to 5 hour
period. The magnitude of the standard deviation of the residuals from the linear fit to the data are considered to be the
time averaged radio core flux density at 8.3 GHz (flat spectrum: $\alpha=0$ is assumed). The second column of Table 2, $n_{rms}$, lists the estimated values
of $n$ that are based on this method. The average of our 8 estimated values is $n_{rms}=2.60 $. For the epochs
50590.16, 51270.33, 51535.63 there is not enough data sampling to use this method and the sample average
is used as a default value.
\par A second method uses the spectral index of the optically thin emission at late times in the
flare, $\alpha$. The late time aspect is important because it allows
sufficient time for the large instabilities in the core flux (corona
and disk) to decay and strong shocks in the ejected plasmoid to be
damped as well. The inherently longer time scales facilitate
discrete time averaging that also diminishes said effects. An important
aspect that was discovered in \citet{pun12} that holds true for the
models presented here is that the leptons responsible for the 2.3
GHz to 8.3 GHz emission are at fairly low energy for relativistic
astrophysical ejecta (with $E$ equal to a few tens at most).
The important consequence is that there is minimal spectral ageing
due to synchrotron losses in the time frames considered, thus the
spectral index is indicative of $n$ for low energy leptons (where
most of the energy resides), even at earlier times. Thus, estimating
$\alpha$ in the optically thin emission at late times is both a critical
and relevant aspect of the calculation.
\par Ideally, we would want VLBA maps of the core at both
frequencies at the time of the estimate of $\alpha$. Unfortunately,
we only have this for one flare, that of 51336.970, as shown in
Figure 13. The next best option is to fit the data at late times for
2.3 GHz, 8.3 GHz and 15 GHz with a two component model. The core
component in the model is chosen to have a flat spectrum $\alpha
\approx 0$ from 2.3 GHz to 15 GHz and one then solves for the
spectral index of the optically thin flux component and the
normalization of the two power law components. The three flux
density data points and the three unknowns provide a unique
solution. For strong major flares, the late time behavior, one to
three days after launch, make the method viable since the optically
thin flux level is well above the random optically thick background
of the time variable core and emission from strong
shocks\footnote{The weak major flares described above are defined
by $\sim$ 100 - 200 mJy of peak optically thin flux density at 2.3
GHz. The stronger major flares are those that are defined to be
above this level and can achieve optically thin flux densities
exceeding 1500 mJy \citep{rod95}. The strong major flares are of
sufficient strength that they are clearly distinguished from other
background radio emission. In the following, the word "major" will
often be dropped in the flare designations of weak and strong.}. The
weak major flares, 50590.16, 51270.33 can not be extricated from the
background fluctuations and the sample average of $n_{2} =2.80$ is
used in Table 2. The same value is chosen for 51535.63 for which
there is insufficient data to make an estimate.
\par The $n$ values that are estimated from the spectral method above
are listed as $n_{2}$ in column 3 of Table 2. The range of $n$ in columns 2 and
3 can be used as an estimate of our uncertainty in $n$ in the following.
The first column of Table 2 is the estimated flare initiation date. The last two columns are the estimated energy within
the ejecta, $E_{\mathrm{plasmoid}}$, and the time averaged power required to launch the ejecta
($Q$), respectively.
\subsection{Doppler Factor Considerations} The solutions depend strongly on the distance to the source $D$ through its
role in the determination of $\delta$ (see the entries in Table 2).
The intrinsic kinematics of apparent superluminal motion are a very
strong function of $D$ \citep{fen99}. Thus it was important to solve
for the plasmoid energy in a plausible range of $D$ and $\delta$.
In order to estimate the Doppler factors we assume that the kinematic
results from \citet{fen99} are common to the entire
time frame from 1997 to 2000 as evidenced by interferometric
observations of multiple flares \citep{dha00,mil05}. The values of
$\delta$ for different values of $D$ are listed in Table 2 and the
model based estimates of the energy and $Q$ were computed
accordingly. From equation (16), the \emph{intrinsic} spectral luminosity associated with an
observed flux density $\sim \delta^{-(3+\alpha)}$. For example, if the
distance to the source is 10.5 kpc instead of 11 kpc, the intrinsic
spectral luminosity changes by a factor
$\sim(0.31/0.54)^{3.9}=(1/8.7)$. Thus, the strong dependence of the
energy on $\delta$ and $D$ that is depicted in Table 2 - a change in
distance from 11 kpc to 10.5 kpc reduces the energy of the plasmoid
by a factor of 5 - 6.

\begin{figure}
\begin{center}
\includegraphics[width= 90 mm, angle= 0]{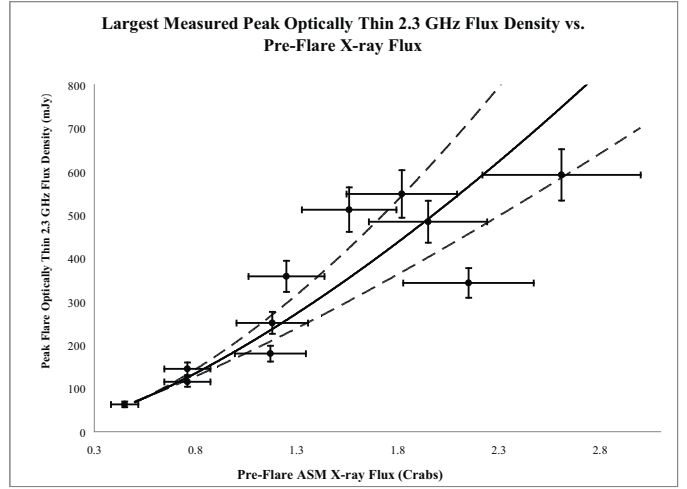}
\caption{\footnotesize{A weighted least squares powerlaw fit (solid
black curve) to the X-ray fluxes 0 to 4 hours before the flares (in
Crabs) and the peak optically thin measured 2.3 GHz flux densities
(mJy). The dashed curves indicate the standard error in the fit from
the first row of Table 5. The errors on the flux densities are
simply the uncertainty in the GBI measurements are from Table 1.}}
\end{center}
\end{figure}

\subsection{Explicit Computation and Comparison to Other Minimum Energy Estimates}
This section is a discussion of the nuances of this method as
compared to conventional minimum energy estimates. This highlights
some of the advantages of the method and illustrates the details of
how a calculation is performed in practice. An instructive example is the
flare at epoch 50750.50 for which a standard minimum energy estimate was
performed with an assumed distance 11 kpc in \citet{fen99}.
\par The calculational method described above has two assumptions. One is that all major flare ejections are considered
to have approximately the same Doppler factor (and we explore a
range of possible distance dependent fixed Doppler factors in Table
2). Secondly, the plasmoids are of similar origin and composition as
those in December 1993 and they will evolve similarly. Thus, the
plasmoid will be near peak optically thin flux density at 2.3 GHz
when it evolves toward $\tau =0.1$ and a minimum energy
configuration. The system is self consistently solved for with
equations (7) to (18). Given our two assumptions, the uncertainty in
our estimation of the energy in the plasmoid is contained in two
quantities, $n$ and $S_{\nu=2.3\mathrm{GHz}}(\tau =0.1)$. The range
of plausible values of $S_{\nu=2.3\mathrm{GHz}}(\tau =0.1)$ are
determined by the light curve in Figure 2 and are listed in column 5
of Table 1. To quantify an uncertainty in $n$ we use the two methods
described in section 3.2 (listed in columns (2) and (3) of Table 2)
to create a plausible range. First, we consider the RMS variations
from a linear fit to the data after the rise of the flare has
terminated. We consider the 8.3 GHz data from the GBI observing run
that started at 50750.83 and ended at 50751.07. A linear fit of the
flux density in mJy versus date yields an intercept of 7874.645 and
a slope of -0.15516. The data was then subtracted from the linear
fit. The standard deviation of the residuals was computed which
yielded an RMS variation of 11 mJy\footnote{As a consistency check,
the same calculation applied to the 15 GHz data a few hours earlier
(the strong oscillations in Figure 2) yields a time averaged core
flux density of 15 mJy}. This was equated to the time averaged
amplitude of an oscillating core flux density. The core was assumed
to be flat spectrum, $\alpha=0$, from 2.3 GHz to 8.3 GHz. The result
was subtracted from the total flux density to yield the optically
thin contribution to the ejection at each epoch of observation. The
two point spectral indices of the plasmoid from 2.3 GHz to 8.3 GHz
were then averaged over the observation, yielding $\alpha=0.573$ for the optically thin component.
Bearing in mind that the points sampled overlap the time frame for
which $\tau =0.1$, we corrected the flux for SSA to get the
intrinsic spectral index of 0.65. The corresponding $n_{rms}=2.30$
is listed in column 2 of Table 2.
\par The second estimate of $n$ from Table 2, $n_{2}$, requires estimates
of the core, weak compact ejecta and optically thick shock emission
at late times in the flare evolution. To determine these
contributions, we average the flux densities at both 2.3 GHz and 8.3
GHz from 50754.825 to 50755.023 (12 points). The 15 GHz Ryle data
slightly precedes the GBI data and we average this flux density from
50754.791 to 50754.808. The time averaged flux densities are 148
mJy, 54 mJy and 35 mJy at 2.3 GHz, 8.3 GHz, and 15 GHz,
respectively. Our two component model yields $\alpha=0.84$ or
$n=2.68$ for the optically thin component and a background flat
spectrum flux density of 6 mJy with $\alpha=0$.
\par The errors in the parameters needed for our calculation are defined by
\begin{eqnarray}
&& 0.65 \leq n \leq 0.84 \;, \\
&& 431 \mathrm{mJy} \leq  S_{\nu=2.3\mathrm{GHz}}(\tau =0.1) \leq
548 \mathrm{mJy} \;.
\end{eqnarray}
\par From the range of input parameters given by Equations (19) and (20), we compute the minimum energy state for $\tau=0.1$ for the four configurations
at the vertices of the rectangular range of 2-D parameter space
defined by $n$ and $S_{\nu=2.3\mathrm{GHz}}(\tau =0.1)$:
\begin{widetext}
\begin{eqnarray}
&& E_{\mathrm{plasmoid}}[n=0.65,S_{\nu=2.3\mathrm{GHz}}(\tau =0.1) =431 \mathrm{mJy}] = 2.13 \times 10^{42} \mathrm{erg/s} \;, \\
&& E_{\mathrm{plasmoid}}[n=0.65,S_{\nu=2.3\mathrm{GHz}}(\tau =0.1) =548 \mathrm{mJy}] = 2.84 \times 10^{42} \mathrm{erg/s} \;, \\
&& E_{\mathrm{plasmoid}}[n=0.84,S_{\nu=2.3\mathrm{GHz}}(\tau =0.1) =431 \mathrm{mJy}] = 4.05 \times 10^{42} \mathrm{erg/s} \;, \\
&& E_{\mathrm{plasmoid}}[n=0.84,S_{\nu=2.3\mathrm{GHz}}(\tau =0.1) =
548 \mathrm{mJy}] = 5.40 \times 10^{42} \mathrm{erg/s} \;.
\end{eqnarray}
\end{widetext}
The estimated energy and its error are then given by the mean and
standard deviation of the numbers above, $E = 3.61 \times 10^{42}
\pm 1.42 \times 10^{42}$ ergs.
\par In order to calculate the power required to eject the plasma we write
\begin{equation}
Q = \Gamma_{rel}E_{\mathrm{plasmoid}}/(T_{end} -T_{start})\;,
\end{equation}
where the relativistic Lorentz factor at 11 kpc is given by
\citet{fen99} as $\Gamma_{rel}=5$. $T_{start}$ and $T_{end}$ are the
plasma ejection start and stop times from Table 1,
\begin{equation}
T_{start} =50750.52\pm 0.02 \quad , \quad T_{end} = 50750.695 \pm 0.055 \;.
\end{equation}
Thus, from Equation (25)
\begin{widetext}
\begin{eqnarray}
&& Q = 5 (3.61 \times 10^{42} \mathrm{ergs})/(0.175\,
\mathrm{days})\pm Q\sqrt{\left(\frac{\delta
E_{\mathrm{plasmoid}}}{E_{\mathrm{plasmoid}}}\right)^{2} +
\left(\frac{\delta T_{end}}{T_{end} - T_{start}}\right)^{2}
+\left(\frac{\delta T_{start}}{T_{end} -  T_{start}}\right)^{2}}\;,
\end{eqnarray}
\end{widetext}
where the error is from the quadrature propagation of error: $\delta
E_{\mathrm{plasmoid}}$, $\delta T_{end}$, $\delta T_{start}$ are the
errors in the energy (above), start and end times, respectively.
From Section 2.1, a minimum error that can be assigned to both $\delta T_{start}$ and $\delta T_{end}$ is
$0.1(T_{end} -  T_{start})$ and this error is added in quadrature to
any other uncertainty to the flare initiation time and end times that are indicated in Equation (26).
In this particular case, equations (24) - (27) yield $Q = 1.19 \times 10^{39} \pm 6.29
\times 10^{38}$ ergs/s, one of the largest relative errors in the
sample. The uncertainty induced by the uncertainty in the distance
to the source (as imprinted through the Doppler factor) is accounted
for by the three rows for each for each flare in Table 2,
corresponding to distances of 10 kpc, 10.5 kpc and 11 kpc (Doppler
factors 0.31, 0.54 and 0.69, respectively).

We highlight some key differences in this estimation technique and a standard minimum energy estimate
in \citet{fen99}.
\begin{itemize}
\item It is noted in \citet{fen99} that the plasmoid size is uncertain and
it was assumed that the radius of the plasmoid is $6.5 \times 10^{14}$ cm. By contrast, this
calculation has one assumption: the plasmoids are of similar
origin and composition as those in December 1993 and they will
evolve similarly. Thus, the plasmoid will be near peak optically
thin flux density at 2.3 GHz when it evolves toward $\tau =0.1$ and
a minimum energy configuration. The system is self consistently
solved for with equations (7) to (18). The uncertainties in our
estimated $n$ and $S_{\nu=2.3\mathrm{GHz}}(\tau =0.1)$ lead to a
range of calculated plasmoid radii $1.75 \times 10^{14}$cm $<R< 1.95 \times
10^{14}$ cm that are derived in the self-consistent solutions
that are defined by equations (7) - (18). The stored energy in the plasmoid
$\sim R^{9/7}$ \citep{fen99}.  Thus, the ratio of the two energy estimates is $\approx (6.5/1.85)^{9/7} = 5.1 $. This
explains the fact that \citet{fen99} computes an energy of $2 \times
10^{43}$ ergs and in Table 2 we find $(3.61 \pm 1.42) \times
10^{42}$ ergs. In relation to this, note that the range of
$S_{\nu=2.3\mathrm{GHz}}(\tau =0.1)$ in Equation (20) and the light
curve in Figure 2 indicate that the calculation is performed in the
interval 50750.83 and 50751.01. From equation (26) this indicates a
time frame for plasmoid expansion of 26000 sec  to 44000 sec in our
models. This equates to a lateral expansion velocity between 0.13c
and 0.25c. By contrast there is an implicit $\approx$ c expansion
speed in the \citet{fen99} analysis since the flux density that they use
in the calculation (550 mJy) is at epoch 50750.86 in combination with their aforementioned
radius. The MERLIN observations in \citet{fen99} were interpreted as
providing an upper limit on the expansion speed of 0.14c. VLA and
VLBA monitoring have been used to estimate an expansion speed of
$\approx$ 0.2c \citep{dha00}. Thus, the self-consistent time
evolutionary model presented here is the more consistent with
observation.
\item \citet{fen99} choose a plasma injection episode of 12 hours to convert $E_{\mathrm{plasmoid}}$ to $Q$. Our analysis
of the plasma start and stop times indicate an energy injection time of $\approx$ 4 hours.
\item The power required to eject the plasmoid in \citet{fen99} is $2\times 10^{39}$ ergs/sec
compared to $Q = 1.19 \times 10^{39} \pm 6.29 \times 10^{38}$ ergs/s
found above. There are two competing effects. The \citet{fen99} estimate is a factor of $\approx 5$
larger for the large assumed radius in the first item above, and a factor of $\approx 3$ smaller for
longer rise time (second item) for a net $\approx$ (5/3) increase from our
estimate.
\end{itemize}
This example was perhaps the most complicated case considered, The same
techniques for estimation of quantities and errors are applied
to the other (simpler) energy and $Q$ estimates in columns 6 and 7 of Table 2,
respectively.

\begin{table*}
\caption{Estimates of Power Required to Eject Major Flares}
{\tiny\begin{tabular}{ccccccc} \tableline\rule{0mm}{3mm}
Estimated Flare &  $n_{rms}$ & $n_{2}$ & D \tablenotemark{b}  & $\delta$ \tablenotemark{c}   & $E_{\mathrm{plasmoid}}$ & $Q$ \\
  Start (MJD) & &   & kpc & & $10^{42}$ ergs & $10^{38}$ ergs/s\\
\tableline \rule{0mm}{3mm}
50590.160             & 2.60 &  2.80          & 11 &  0.31  & $0.32 \pm 0.08$  &  $4.55 \pm 1.27$\\
50590.160             & 2.60 & 2.80           & 10.5 &  0.54  & $0.066 \pm 0.017$  &  $0.60 \pm 0.16$\\
50590.160             & 2.60  & 2.80           & 10 &  0.69  & $0.031 \pm 0.008$  &  $0.22 \pm 0.06$\\
50750.520              & 2.30  & 2.68           & 11 &  0.31  & $3.61 \pm 1.42$  &  $12.80 \pm 6.29 $\\
50750.520              & 2.30  & 2.68           & 10.5 &  0.54  & $0.78 \pm 0.28$  &  $1.55\pm 0.83$\\
50750.520              & 2.30  & 2.68           & 10 &  0.69  & $0.37 \pm 0.13$  &  $0.57 \pm 0.29$\\
50916.140             & 2.78  & 2.92           & 11 &  0.31  & $11.70 \pm 3.07$  &  $19.34 \pm 5.71$\\
50916.140             & 2.78  & 2.92             & 10.5 &  0.54  & $2.05\pm 0.62$  &  $2.10 \pm 0.71$\\
50916.140             & 2.78  & 2.92            & 10 &  0.69  & $1.00 \pm 0.27 $  &  $0.80 \pm 0.24$\\
50967.200             & 2.40  & 2.80             & 11 &  0.31  & $4.70 \pm 2.97$  &  $20.9 \pm 13.6$\\
50967.200             & 2.40  & 2.80      & 10.5 &  0.54  & $0.99 \pm 0.63$  &  $2.73 \pm 1.78$\\
50967.200             & 2.40  & 2.80      & 10 &  0.69  & $0.47 \pm 0.30$  &  $0.99 \pm 0.65$\\
51003.070             & 2.76  & 2.84             & 11 &  0.31  & $2.60 \pm 0.31$  &  $9.01 \pm 1.69$\\
51003.070             & 2.76  & 2.84     & 10.5 &  0.54  & $0.51 \pm 0.07$  &  $1.10 \pm 0.22$\\
51003.070             & 2.76  & 2.84     & 10 &  0.69  & $0.24 \pm 0.03$  &  $0.40 \pm 0.08$\\
51270.330             & 2.60  & 2.80             & 11 &  0.31  & $1.07 \pm 0.24$  &  $5.16 \pm 1.82$\\
51270.330             & 2.60  & 2.80     & 10.5 &  0.54  & $0.24 \pm 0.06$  &  $0.73 \pm 0.29$\\
51270.330             & 2.60  & 2.80     & 10 &  0.69  & $0.11 \pm 0.03$  &  $0.26 \pm 0.10$\\
51336.970             & 2.74  & 2.88\tablenotemark{a}              & 11 &  0.31  & $5.79 \pm 0.97$  &  $16.34 \pm 4.24$\\
51336.970             & 2.74  & 2.88\tablenotemark{a}      & 10.5 &  0.54  & $1.22 \pm 0.20 $  &  $ 2.13 \pm 0.55$\\
51336.970             & 2.74  & 2.8 \tablenotemark{a}      & 10 &  0.69  & $0.56 \pm 0.08$  &  $0.76 \pm 0.18$\\
51375.150             & 2.62  & 2.96          & 11 &  0.31  & $1.81 \pm 0.74$  &  $8.06 \pm 3.44$\\
51375.150             & 2.62  & 2.96     & 10.5 &  0.54  & $0.38 \pm 0.15 $  &  $1.06 \pm 0.43$\\
51375.150             & 2.62  & 2.96     & 10 &  0.69  & $0.18 \pm 0.07$  &  $0.38 \pm 0.15$\\
51499.702             & 2.62  & 2.76             & 11 &  0.31  & $3.22 \pm 0.71$  &  $10.77 \pm 2.60$\\
51499.702             & 2.62  & 2.76           & 10.5 &  0.54  & $0.89 \pm 0.20$  &  $1.86 \pm 0.44$\\
51499.702             & 2.62  & 2.76     & 10 &  0.69  & $0.38 \pm 0.09$  &  $0.61 \pm 0.15$\\
51535.630             & 2.60  & 2.80          & 11 &  0.31  & $5.23 \pm 1.58$  &  $15.13 \pm 4.86$\\
51535.630             & 2.60  & 2.80     & 10.5 &  0.54  & $1.11 \pm 0.33$  &  $1.99 \pm 0.63$\\
51535.630             & 2.60  & 2.80     & 10 &  0.69  & $0.52 \pm 0.16$  &  $0.73 \pm 0.23$\\
51602.506             & 2.70  & 2.90             & 11 &  0.31  & $1.10 \pm 0.23$  &  $5.07 \pm 1.18$\\
51602.506             & 2.70  & 2.90              & 10.5 &  0.54  & $0.23 \pm 0.04$  &  $0.65 \pm 0.14$\\
51602.506             & 2.70  & 2.90             & 10 &  0.69  & $0.11 \pm 0.02$  &  $0.24 \pm 0.05$\\
\end{tabular}}
\tablenotetext{a}{There was a simultaneous a 2.3 GHz VLBA
observation and a 8.3 GHz observation (see Figure 13). The core flux
density was 21 mJy and 42 mJy at 2.3 GHz and 8.3 GHz, respectively
(Vivek Dhawan private communication, 2012). Simultaneous GBI data
indicated flux densities of 378 mJy and 146 mJy at 2.3 GHz and 8.3
GHz, respectively on 51338.23. Subtracting off the core flux density
yields $\alpha=0.94$ or $n=2.88$.} \tablenotetext{b}{distance to
source} \tablenotetext{c}{Doppler factor compatible with kinematics
and the distance to source in the previous column}

\end{table*}

\section{RXTE X-ray Flux} In order to make a connection between the
accretion flow state before and during the major ejections,
one needs X-ray observations that cover a wide range of energies as
there is a significant contribution to the intrinsic luminosity from the black body
energies of $\sim 1$ keV and a corona with significant luminosity up
to $\sim$ 50 keV or more \citep{don04,fuc03,rod09}. Apriori, one
cannot assume that one particular range of energies is more connected
to the production of major flares than another as the physics of the
accretion flow and flare production are basically unknown. Thus, the
broadest possible X-ray coverage is required to make any significant
conclusions. As discussed in the Introduction, the ejections occur
unexpectedly on the order of hours so there is no time for target of
opportunity X-ray observations just before the flare starts and
during the rise, all data is serendipitous. Thus, there must be the
densest possible survey time sampling in combination with fortunate
circumstances. During the period that was considered, the only viable alternative is the All Sky Monitor
(ASM) of the Rossi X-ray Timing Explorer (RXTE). The data is not
sufficient to produce spectra and only covers a very narrow range of
energies, 1.2 keV to 12 keV. The ASM light curves are provided in
counts/sec so it would be advantageous to
\begin{enumerate}
\item convert count rates to fluxes
\item convert observed flux to intrinsic (unabsorbed) flux
\item extend the estimated measured flux from the range 1.2 keV to
12 kev to 1.2 keV to 50 keV
\item find methods for ameliorating systematic errors in the ASM
count rates in individual bins in the data that is provided by the
ASM team.
\end{enumerate}
The final result is an estimator that takes the counts measured in
the three bins of the ASM data, 1.2--3~keV, 3--5~keV and 5--12~keV, and converts this to an
intrinsic (unabsorbed) X-ray luminosity from 1.2--50~keV. Our calibrations indicate a final result with an intrinsic
error of less than 15\%, assuming a constant absorption column
density.
\subsection{The ASM Data Measured in Crabs} The first step in our
process was to convert the ASM count rates to Crab units and use the
archival Crab spectra to later convert this to fluxes. This initial
effort had some startling results which motivated the in-depth
analysis and modeling that appeared in the previous section and the
rest of this paper. We explored two methods of converting counts per second to
Crabs. Firstly, we converted the counts/second (cts/s) from
1.2--12~keV to Crabs: 75.41 cts/s in the ASM band from 1.2--12~keV
is the equivalent of 1 Crab. This conversion is based on the 16
year archival average of ASM data. But, since the Crab spectra are
harder than the GRS~1915+105 spectra, it was more accurate in our
final conversion to cgs flux units to convert the cts/s in each bin
to Crabs separately then convert to a flux in each bin. The first
method of using the whole ASM band overestimated the flux. We define
the three ASM bins, 1.2--3~keV, 3--5~keV and 5--12~keV, as bin 1,
bin 2 and bin 3, respectively. The counts rates in each bin are
defined symbolically as
\begin{mathletters}
\begin{eqnarray}
&& C1 \equiv \mathrm{cts/s \; in \; bin\;1} \;, \\
&& C2 \equiv
\mathrm{max}(5.75, \mathrm{cts/s \; in \; bin\;2})\;,\\
&& C3 \equiv \mathrm{cts/s \; in \; bin\;3}\;.
\end{eqnarray}
\end{mathletters}
The lower limit in the expression for C2 arises from a systematic
error that occurs in bin 2 sporadically. Sometimes at low count
rates there is a "drop out" in this channel and count rates far
below the much more accurate PCA archival count rate for this energy
range are registered. This is clearly a systematic ASM error and the
5.75 cts/s are the equivalent of the lowest archival recorded PCA
count rate based on a comparison between PCA and ASM using the flux
conversion for ASM in equation (30b), below. This minimum cutoff is
derived and discussed in detail in Sections 4.2.1 and 4.2.4.
\par The archival ASM Crab count rates are very stable in each
bin with only a few percent variation over the epochs that are
considered in this treatment. Thus, we used the archival average
count rates to convert the count rates in the three bins to fluxes
in Crabs , $Fi(\mathrm{Crab})$, $i$ = 1, 2, 3, by the equations
\begin{mathletters}
\begin{eqnarray}
&& F1(\mathrm{Crab}) =C1/26.78 \;, \\
&& F2(\mathrm{Crab}) =C2/23.25\;,\\
&& F3(\mathrm{Crab}) =C3/25.38\;.
\end{eqnarray}
\end{mathletters}
In order to compute the errors in each bin, we combined the errors
for each of the GRS~1915+105 count rates with the mean error on the
Crab count rate, and the variance of the Crab rate in each bin. We
then propagated the errors as if they were independent. This
independence is not formally true, but we consider this of small
consequence since the variance in each bin is not that high, and the
statistical errors on the rates are not that high either.
\par Figure 14 shows that the peak measured optically thin flux density
at 2.3 GHz (column 6 of Table 1) is strongly correlated with the ASM
X-ray flux in Crabs (computed from the entire ASM band). The Pearson
(Spearman Rank) probability of the scatter occurring by random chance is $1.9
\times 10^{-5}$ ($5.2 \times 10^{-3}$). A weighted least squares
powerlaw fit (with an index of $1.45 \pm 0.17$) is very good with
only the one outlier, 50967.20, with the aforementioned large gap
in GBI coverage near the peak of the optically thin flux density at
2.3 GHz \citep{ree89}. The fit improves somewhat if we use the
estimated value of $S_{\nu=2.3\mathrm{GHz}}(\tau =0.1)$ from Table 1
since this is effectively an average flux density near the peak flux
density and is less sensitive to gaps in the observations. The
improved weighted least squares powerlaw fit has an index of $1.40
\pm 0.11$. The probability that the data scatter occurs by random
chance and the two quantities are not correlated is 0.26\% according
to a Spearman rank correlation test.
\par Figure 14 represents an empirical relationship
between the X-ray flux 0 to 4 hours before the ejection of a
flare and the peak optically thin low frequency flux of the flare.
It is model independent. We believe this to be a fundamental
property of the physics of major flare launching and it motivates
the detailed model dependent analyses that follow. In it most
transparent guise, these correlations seem to indicate a direct
relationship between the luminosity of the accretion flow just prior
to major flare ejection with the energy of the flare and/or the
power required to launch the flare.
\begin{figure}
\begin{center}
\includegraphics[width=90 mm, angle= 0]{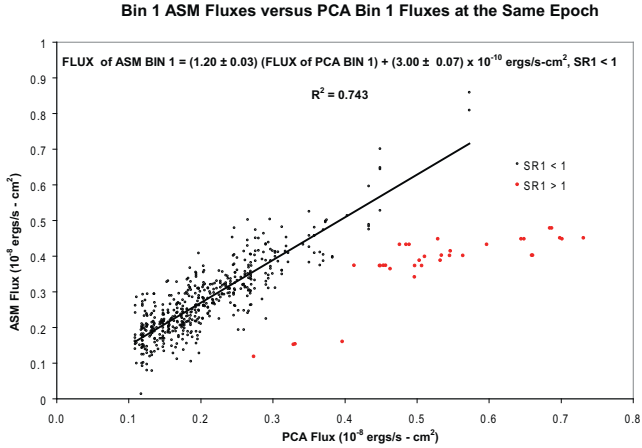}
\caption{\footnotesize{A comparison of bin 1 fluxes in
quasi-simultaneous observations of ASM and PCA. The units on the
axes are $10^{-8}$ ergs/sec/$\mathrm{cm}^{2}$. $R^{2}$ is the
squared multiple regression coefficient.}}
\end{center}
\end{figure}
\subsection{Converting ASM Bin Counts to X-ray Luminosity} Thusly
motivated, we develop the physical interpretation of the ASM data.
The physical meaning of Figure 14 would be greatly enhanced
if the ASM fluxes in Crabs could be related quantitatively to the
broadband energy released as X-rays by the accretion flow. This
effort is challenged by the extrinsic and intrinsic sources of
absorption and the narrow range of energies detected by ASM. This is
a large extrapolation of the data and its success is not guaranteed
ahead of time. We show that for a wide range of X-ray spectral
states that estimators with an accuracy on the order of 15\% are
achieved. This endeavor requires that all the information in the ASM bins
must be used and the results must be verified by comparison to a
large database of quality models of quasi-simultaneous broadband PCA
observations. The ASM data not only give fluxes, but the ratios of
the fluxes in the various bands contain information on the
spectral slope that can be used to extrapolate the energies higher than
those that are observed.
\subsubsection{Converting ASM Crab Fluxes to cgs Units} As a first step in this process,
it is useful to define softness ratios in terms of fluxes in the
three ASM bins \footnote{We call these softness ratios so they are
not confused with the various definitions in the literature of flux
ratios that are designated as hardness ratios}. The first step is to
convert Crab fluxes to cgs fluxes. To do so, we downloaded, reduced
and analyzed all  RXTE/PCA observations of the Crab nebula performed between 1996 February, 1 (MJD 50114) and 2011
December, 31 (MJD 55926).

\par The RXTE/PCA data were reduced following standard steps  (e.g.
\citep{rod09}) with the {HEASOFT} v6.12 to obtain spectra and
response matrices. GTI were defined following the recommended filter
criteria
\footnote{e.g. http://heasarc.gsfc.nasa.gov/docs/xte/pca\_news.html\#practices}.
We then extracted source and background spectra from the top layer
of PCA/Proportional Counter Unit \#2, using the bright source background model,
while the response matrix was produced with accordance to the gain
epochs of the PCA. The resultant spectral products were loaded into {\tt XSPEC} v12.7.1 and fitted. \\
\indent As we wanted to obtain fluxes in well defined bands, we
fitted the spectra with a simple power law model, only ensuring that
the reduced $\chi^2$ was less than 1.8 for the fit (and fluxes) to
be acceptable.  This resulted in about 500 'good' fits. As the Crab
spectrum was fairly constant during the epochs spanned by the flares
in Table 1 (most of the apparent changes over time were due to epoch
changes and compensated for by taking the appropriate matrices), the
fluxes in the 1.5--3~keV, 3--5~keV, 5--12~keV, 12--50~keV were then
obtained by taking the mean over the whole period of observations.
\begin{widetext}
\begin{mathletters}
\begin{eqnarray}
&& F1=F1(\mathrm{Crab})(1.05 \times 10^{-8})\mathrm{ergs/s-cm^2}= C1(3.92 \times 10^{-10})\mathrm{ergs/s-cm^2} \;, \\
&& F2=F2(\mathrm{Crab})(0.731 \times 10^{-8})\mathrm{ergs/s-cm^2}= C2(3.14 \times 10^{-10})\mathrm{ergs/s-cm^2}\;,\\
&& F3=F3(\mathrm{Crab})(1.17 \times 10^{-8})\mathrm{ergs/s-cm^2}= C3(4.61 \times 10^{-10})\mathrm{ergs/s-cm^2}\;,
\end{eqnarray}
\end{mathletters}
\end{widetext}
where the right hand side of the expressions is derived from Equation (29).
In regards to the definition in equation (28b) for C2, the minimum
observed PCA flux between 3 keV and 5 keV is $\approx 1.8 \times
10^{-9}\mathrm{ergs/s-cm^2}$ except for a small handful of anomalous
outliers that could be artifacts of observational errors or data
reduction. Equation (30) represents the flux calibration of the ASM count rates.
\subsubsection{The Database of PCA Spectra Used to Calibrate the ASM Estimator of Luminosity}
It is desirable to convert the observed fluxes in the three bins to
a broadband (unabsorbed) intrinsic flux from 1.2--50~keV which is directly
representative of the physical state of the accretion flow of GRS
1915+105. This requires extrapolating the flux beyond 12 keV to 50
keV and de-reddening the fluxes. The program for accomplishing this
will be based on comparisons of ASM observations with PCA
observations that are simultaneous to within 45 minutes. The PCA
data has sufficient spectral coverage and sensitivity that model
fits to the data can be performed that allow us to estimate the
absorption on the line of sight for the various bins and detailed
intrinsic spectral parameters. In order to permit a high fidelity
comparison of the PCA and the ASM data that are asynchronous by as
much as 45 minutes requires that the PCA fluxes of the chosen epochs be stable over similar time
scales. This will also permit higher signal to noise spectra.
\par In \citet{bel00}, the spectral states of GRS~1915+105 were categorized in 12 classes.
This is based on the PCA light curves and color-color diagrams. From
this classification one can see that there are 2 stable classes on
time scales $>1$ second. These are the $\chi$ and $\phi$ classes.
These more or less correspond to the two stable sets of hard and
soft states that we found in the PCA data for GRS~1915+105 that were
suitable for modeling because of their stability. In terms of ASM
data these PCA selected states also turn out to be distinct. To see
this, consider, the notion of softness ratios defined from equation
(31)
\begin{mathletters}
\begin{eqnarray}
&& SR1=F1/F2=0.93(C1/C2) \;, \\
&& SR2=F2/F3=0.68(C2/C3)\;.
\end{eqnarray}
\end{mathletters}
The soft states lie at SR2 values above the line
\begin{equation}
SR2 = 1.15 (SR1) + 0.28 \;,
\end{equation}
in the SR1-SR2 plane and the hard states lie at SR2 values below this line. The
concept of bin based fluxes that are used in equations (31) and (32)
apply to PCA data if we make designations analogous to equation
(29),
\begin{mathletters}
\begin{eqnarray}
&& F1(PCA) \equiv \mathrm{flux \; in \; bin\;1} \;, \\
&& F2(PCA) \equiv \mathrm{flux \; in \; bin\;2}\;,\\
&& F3(PCA) \equiv \mathrm{flux \; in \; bin\;3}\;,\\
&& F4(PCA) \equiv \mathrm{flux \; in \; bin\;4}\;.
\end{eqnarray}
\end{mathletters}
Where PCA(bin 2) and PCA(bin 3) are the same as for ASM, 3--5~keV
and 5--12~keV, respectively. PCA(bin1) is narrower than for
ASM, 1.5--3~keV. PCA(bin4) is not covered by ASM observations,
12--50~keV. We found 630 hard ($\chi$) states and 164 soft
($\phi$) states.

\par The spectra where fitted with a model consisting of (absorbed) thermal Comptonization
and a Gaussian to account for a fluorescent iron line.  N$_H$ was
fixed to $5.7 \times 10^{22}\mathrm{cm}^{2}$, (see
(\citet{bel97,mun99,rod09})), since leaving it free to vary leads in
many occurrences of unphysical values. Good fits (with a reduced
$\chi^2 <1.8)$ was achieved for 501 spectra \footnote{None of the
soft state observation are coincident or close in time to a
superluminal ejection so we do not consider them further here.}.
This does not mean that the other spectra are "bad". This means that
our simple model is not able to approximate, or does not represent,
the spectrum well enough so that we can trust the fluxes obtained
with it. These extra 20\% of the spectra probably need extra
components such as disk and/or additional power laws to achieve a
reliable fit.

\begin{table*}
\caption{X-ray Luminosity Evolution (Based on D = 11 kpc)}
{\tiny\begin{tabular}{cccccccccc} \tableline\rule{0mm}{3mm}
24 Hour &Baseline &  Pre-Flare  & 24 Hour & Baseline  & Pre-Flare & Average Rise  & Baseline  & Pre-Flare & Average Rise \\
Baseline  & Date  &  Date  & Baseline & $L_{\mathrm{5-10 hr}}$ & $L_{\mathrm{pre-flare}}$ & $L_{\mathrm{rise}}$  & 5 - 12 keV & 5 - 12 keV & 5 - 12 keV\\
Date&(MJD)  & (MJD) & $L_{\mathrm{24 hr}}$ &  &  &  & Flux ($10^{-8}$ & Flux ($10^{-8}$ & Flux ($10^{-8}$\\
MJD &  &  & ($10^{39}$ ergs/s)& ($10^{39}$ ergs/s) & ($10^{39}$ ergs/s) & ($10^{39}$ ergs/s) & ergs/s- $\mathrm{cm}^2$) &  ergs/s- $\mathrm{cm}^2$) & ergs/s- $\mathrm{cm}^2$)\\

\tableline \rule{0mm}{3mm}
50589.093 & 50590.090             & 50590.157    & $0.39 \pm 0.05$ & $0.40 \pm 0.05$ & $0.41 \pm 0.05$  & $0.35 \pm 0.05$ \tablenotemark{b} & $ 0.65\pm 0.7 $ & $0.66 \pm 0.06$& $0.57\pm 0.06$\tablenotemark{b}\\
50749.356 & 50750.284             & 50750.351    & $0.50 \pm 0.07$ & $0.84 \pm 0.12$ & $1.05 \pm 0.15$  & $0.71 \pm 0.1 $  & $ 2.25\pm 0.23$ & $2.94 \pm 0.29$ & $ 1.63 \pm 0.16$\\
50915.091 & 50916.024             & 50916.146\tablenotemark{a}       & $0.47 \pm 0.06$ & $0.61 \pm 0.07$ & $1.36 \pm 0.18$   & $1.15 \pm 0.15 $  & $ 1.39 \pm 0.14 $ & $3.91 \pm 0.39$ & $ 3.35 \pm 0.34$\\
50915.091 & 50916.024             & 50916.089\tablenotemark{a}       & $0.47 \pm 0.06$ & $0.61 \pm 0.07$ & $1.28 \pm 0.18$   & $1.17 \pm 0.15 $  & $ 1.39 \pm 0.14 $ & $3.79 \pm 0.38$ & $ 3.35 \pm 0.34$\\
50966.101 & 50966.908             & 50967.148    & $0.41 \pm 0.06$ & $0.50 \pm 0.07$ & $1.19 \pm 0.17$  & $ 1.03 \pm 0.15 $  & $ 1.01 \pm 0.1 $ & $ 3.72 \pm 0.37 $ & $ 3.03 \pm 0.30$\\
51001.912 & 51002.770             & 51002.907    & $0.42 \pm 0.06$ & $0.52 \pm 0.07$ & $0.76 \pm 0.11$  & $ 0.45 \pm 0.06 $  & $ 1.36\pm 0.14$ & $1.94 \pm 0.19 $ & $0.85 \pm 0.09$\\
51269.314 & 51270.250             & 51270.315    & $0.46 \pm 0.06$ & $0.58 \pm 0.07$ & $0.54 \pm 0.07$  & ...  & $ 1.31\pm 0.13 $ & $ 1.21 \pm 0.12 $ & ...\\
51335.942 & 51336.870             & 51336.940    & $0.41 \pm 0.06$ & $0.80 \pm 0.10$ & $1.07 \pm 0.14$  & $0.89 \pm 0.12 $  & $ 1.85 \pm 0.19 $ & $2.88 \pm 0.29$ & $2.54 \pm 0.25$\\
51373.967 & 51374.962             & 51375.099    & $0.43 \pm 0.06$ & $0.69 \pm 0.09$ & $0.71 \pm 0.09$ & $0.55 \pm 0.07 $  & $ 1.54 \pm 0.15 $ & $ 1.81 \pm 0.18$ & $ 1.32 \pm 0.13 $\\
51498.555 & 51499.415             & 51499.614    & $0.52 \pm 0.07$ & $0.57 \pm 0.08$ & $0.79 \pm 0.11$  & ...  & $ 1.22 \pm 0.12 $ & $ 1.72 \pm 0.17 $& ...\\
51534.490 & 51535.422             & 51535.556    & $0.43 \pm 0.06$ & $0.56 \pm 0.08$ & $0.90 \pm 0.13$  & ...  & $ 1.15\pm 0.12 $ & $ 2.33 \pm 0.23 $& ...\\
50610.410 & N/A                   & 51602.361    & $0.50 \pm 0.07$ & ... & $0.60 \pm 0.09$  & ...  & ... & $ 1.15\pm 0.06 $ & ...\\
\end{tabular}}
\tablenotetext{a}{There is very little difference in the measured
fluxes on 50916.146 and 50916.089, so the uncertainty in the flare
initiation time from Table 1 does not create significant uncertainty
in our results.} \tablenotetext{b}{Includes epoch 50590.16 for which
some X-ray observations exist within a range spanned by the 10\%
uncertainty that was assigned to all flare start time estimates in
Section 2.1}
\end{table*}

\subsubsection{Systematic Difference in ASM and PCA Data}
Before we can implement the use of an estimator for the ASM data,
we must understand the differences
between measured ASM and PCA fluxes (derived from count rates for
ASM per equation (29)). We were able to find systematic differences
by finding quasi-simultaneous pairs of ASM and PCA observations for
our sample of 501 hard PCA spectra. We
found 531 pairs of observations that were simultaneous to within 45
minutes. The large number of pairs arises because more than one ASM
observation can be within 45 minutes of a PCA observation. We note
three major issues
\begin{enumerate}
\item Bin 1 for ASM is wider than bin 1 for PCA, 1.2--3~keV  as
opposed to 1.5--3~keV
\item As noted in regards to equations (28) and (29), there is a
"drop out" effect in the ASM data in bin 2 for low count rates that
occurs unpredictably
\item Considering the entire archive of PCA observations, SR1 is
is rarely greater than 1 ($\ll 1\% $) and in these rare cases it is
just slightly above 1. Yet $SR1
> 1$ about 29\% of the time in the ASM database with such large values and
frequency that it is unlikely to be explained solely by items (1)
and (2) above. It is not clear how much item (2) contributes to
this, so we must assume that this is an independent instrumental
effect and a source of uncertainty in the ASM data. We compared
large SR1 epochs from ASM to quasi-simultaneous PCA data taken
$\sim$ 10 - 30 minutes separated in time when the source was in what
appeared to be a steady state. The PCA softness ratios were
generally modest and "typical" for the source.
\end{enumerate}
\par In order to assess item (1) above, we created a scatter plot of
the flux in bin 1 for quasi-simultaneous ASM observation with the
flux in bin 1 for the corresponding PCA observations. The results
are plotted in Figure 15. The fit to the scatter plot only
represents epochs with $SR1<1$ in the ASM data-set. The plot
indicates that assuming 20\% more flux would replicate expanding the
PCA bin 1 from 1.5--3 keV to the range of 1.2--3 keV. This
is the method that we will use to compensate for the bin size
difference in our comparison of PCA and quasi-simultaneous ASM data.

\par Interestingly, the 33 $SR1>1$ epochs have a disjoint distribution
concentrated toward the lower right of the plot in Figure 15. The
implication is that $SR1>1$ ASM measurements are systematically
different than other epochs and need to be treated as such. It is
unclear how much of this difference is due to intrinsically large
soft fluxes (i.e., a very bright disc) during these epochs in
combination with the softer ASM bins or a consequence of the
aforementioned instrumental effect. The vast majority of
observations with $SR1>1$ in ASM are very luminous hard steady
states that are characterized by an elevated soft flux in the
quasi-simultaneous PCA observations (see Figure 15). The flux excess
above the PCA sample average is largest in bin 1 (86\% above the sample average) decreasing with
each more energetic bin, so that the bin 4 fluxes are approximately
the sample average. It turns out that the $SR>1$ states are rather
common just before and during flares (see Table 4), so it is
desirable to be able to estimate the luminosity accurately in this
situation. Not knowing why errors occur in ASM measurements during
these luminous states, we conclude that the best way to deal with
this dichotomy is by treating the systematic differences in $SR1>1$
and $SR1<1$ ASM epochs, empirically; i.e. fit the data separately in
a way that compensates for potential systematic errors in the ASM
data. In practice, this empirical method for dealing with item 3 is
to segregate the $SR1>1$ epochs and fit them to the PCA data as an
independent data set.
\par The method of dealing with item 2 was discussed earlier with
regards to equations (28) and (29). The lowest recorded PCA flux in
bin 2 is $1.8 \times 10^{-9}$ ergs/sec/$\mathrm{cm}^{2}$. This will
be treated as a minimum allowed value for an ASM flux in bin 2. As
such, any ASM measurement lower than this is very uncertain with
only a lower limit of $1.8 \times 10^{-9}$
ergs/sec/$\mathrm{cm}^{2}$ designated. This uncertainty must be
compensated for empirically by the choice of the estimator that
minimizes the error associated with this expedience. In general, we
had sufficient data from observations just before or just after an
observation for which we had to invoke the minimum cutoff in
equation (28b) that afforded us the luxury of being able to exclude
these data from our analysis. The lone exception is epoch 51002.907
since this data (with equation (28b) applied) is the only available data that is close in time to the flare start.

\begin{figure*}
\begin{center}
\includegraphics[width=80 mm, angle= 0]{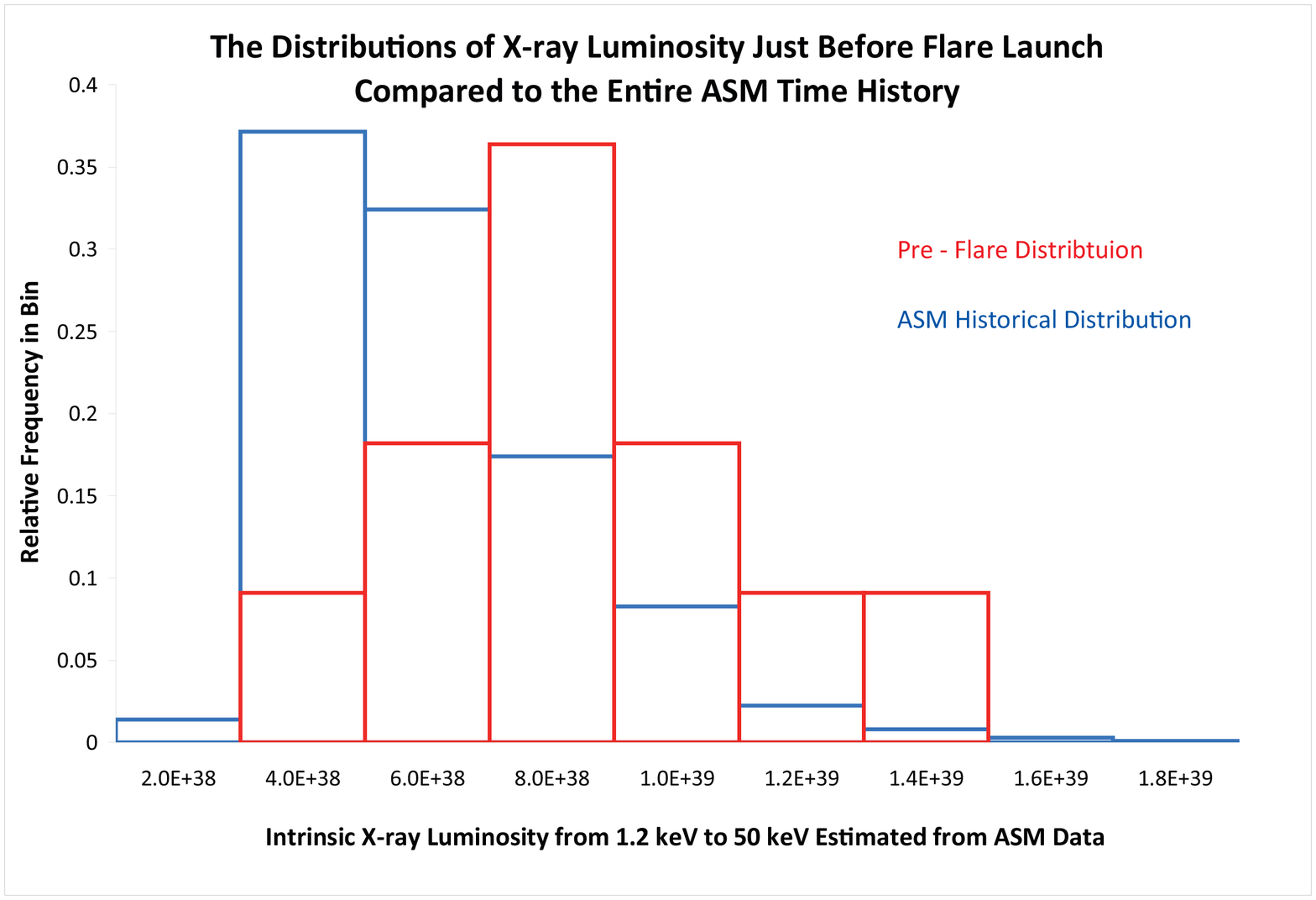}
\includegraphics[width= 80 mm, angle= 0]{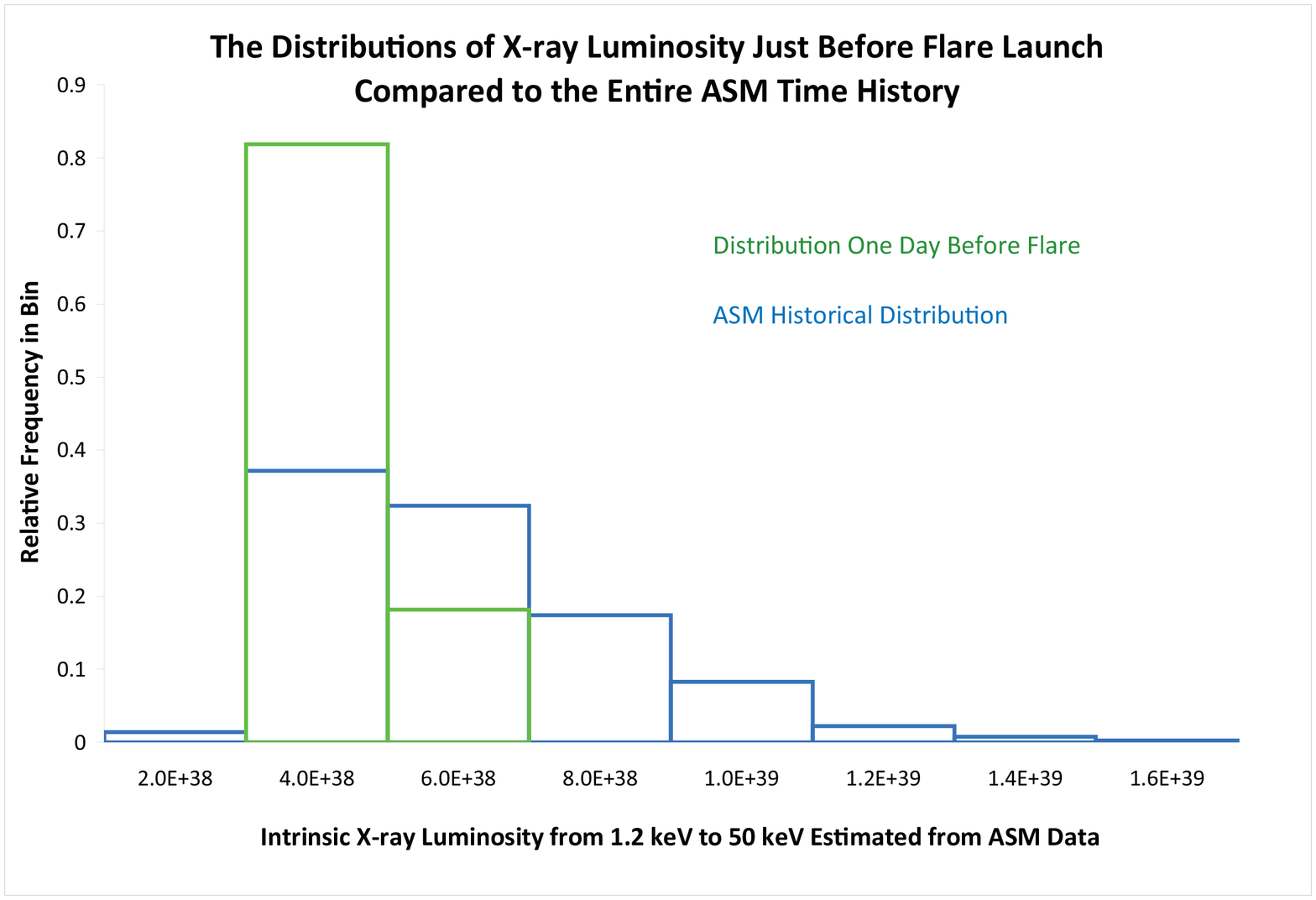}
\caption{\footnotesize{Left panel: the distribution of the intrinsic
X-ray luminosity 0 to 4 hours before a major flare is ejected
compared to the archival distribution of intrinsic X-ray luminosity
that is estimated from the entire ASM archive. Right panel: the
distribution of the intrinsic X-ray luminosity one day before a
major flare is ejected compared to the archival distribution of
intrinsic X-ray luminosity that is estimated from the entire ASM
archive. The luminosity tends to be elevated just prior to the flare
ejection as opposed to being relatively low a day before. The plots
assume D = 11 kpc.}}
\end{center}
\end{figure*}

\begin{figure}
\begin{center}
\includegraphics[width=90 mm, angle= 0]{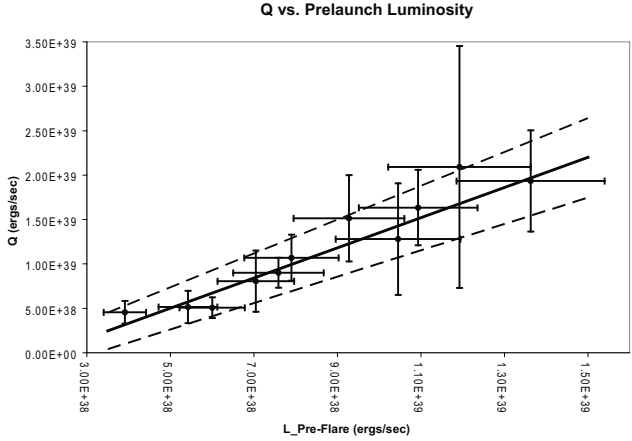}

\caption{\footnotesize{Estimated time average power of the flare
ejection, Q, (Table 2) vs X-ray luminosity 0 to 4 hours before a
major flare ejection (from Table 3). The solid black line is the
weighted least squares fit from Table 5. The dashed black lines
indicate the standard error to the fit from Table 5. A strong
correlation is obvious (see text).}}
\end{center}
\end{figure}
\subsubsection{The ASM Estimator of Intrinsic X-ray Flux}Using the
strategy described in the previous section for dealing with
systematic difference between ASM and PCA bin data, we obtained the
following estimator, $F_{0}$ of $F_{\mathrm{intrinsic}}$ (the
intrinsic, unabsorbed flux derived from PCA models) based on ASM
count rates
\begin{widetext}
\begin{eqnarray}
&&  F_{0}= 0.561\left[(4.363F1(SR1)^{0.2772})
+(1.3767F2(SR1)^{0.0255}) +F3(1+2.730e^{-(2.114SR2)})\right] + 1.25
\times 10^{-8} \mathrm{ergs/s}/\mathrm{cm}^{2} \;, \;
SR1<1\;,\nonumber \\
&&
\end{eqnarray}
\begin{eqnarray}
&& F_{0} = 0.478\left[(4.363F1(SR1)^{0.2772})
+(1.3767F2(SR1)^{0.0255}) +F3(1+2.730e^{-(2.114SR2)})\right] + 1.08
\times 10^{-8} \mathrm{ergs/s}/\mathrm{cm}^{2}\;. \;
SR1>1\;,\nonumber\\
&&
\end{eqnarray}
\end{widetext}
There are many potential sources of errors in the estimator: the
data, the models as well as error induced by variability and time offsets in the ASM and PCA data sampling.
Instead of making assumptions about the manner in which the error propagates, we
empirically determine the error by direct comparison of the estimator
with the flux from the modeled PCA data.
The relative error $\sigma$ for the estimator in equations (34) and (35) is
computed as the standard deviation of the difference, $F_{\mathrm{intrinsic}} - F_{0}$
\begin{eqnarray}
&& \sigma \left(\frac{F_{\mathrm{intrinsic}}
-F_{0}}{F_{\mathrm{intrinsic}}}\right) = 0.131\;, \; SR1<1\;,\\
&& \sigma \left(\frac{F_{\mathrm{intrinsic}}
-F_{0}}{F_{\mathrm{intrinsic}}}\right) = 0.143\;, \; SR1>1\;,
\end{eqnarray}
The empirical fit in equations(34) and (35) has a strange functional form as a consequence of fitting the
four PCA bins in equation (33), separately as a preliminary step. These expressions are not unique nor are they aesthetically beautiful
but they do reproduce the PCA modeled data accurately as evidenced by equations (36) and (37).
\begin{table*}
\caption{X-ray Spectral Evolution } {\tiny\begin{tabular}{cccccccc}
\tableline\rule{0mm}{3mm}
Baseline\tablenotemark{a} &  Pre-Flare  & Baseline\tablenotemark{a}  & Pre-Flare & Average Rise  & Baseline\tablenotemark{a}  & Pre-Flare & Average Rise \\
  Date  &  Date  & SR1 & SR1 & SR1 & SR2 & SR2 & SR2\\
   \tableline \rule{0mm}{3mm}
50590.090            & 50590.157    & 0.61 & 0.63  & 0.54\tablenotemark{b}   & 0.70 & 0.60 & 0.60\tablenotemark{b} \\
50750.284             & 50750.351    & 1.49 & 0.86  & 1.33  & 0.20 & 0.40 & 0.34\\
50916.024             & 50916.146    & 0.66 & 0.54   & 1.77  & 0.52 & 0.54 & 0.20\\
50966.908             & 50967.148    & 0.66 & 0.84  & 1.59  & 0.55 & 0.32 & 0.31\\
51002.770             & 51002.907    & 0.61 & 2.42  & 0.63  & 0.43 & 0.10 & 0.50\\
51270.250             & 51270.315    & 0.64 & 0.61  & ...  & 0.46 & 0.48 & ...\\
51336.870             & 51336.940    & 2.79 & 2.49  & 1.68  & 0.11 & 0.11 & 0.20\\
51374.962             & 51375.099    & 1.61 & 0.62 & 0.54  & 0.31 & 0.39 & 0.47\\
51499.415             & 51499.614    & 0.73 & 2.33  & ...  & 0.49 & 0.2 & ...\\
51535.422             & 51535.556    & 0.79 & 2.48  & ...  & 0.44 & 0.13 & ...\\
N/A                   & 51602.361    & ... & 0.46  & ...  & ... & 0.61 & ...\\
\end{tabular}}
\tablenotetext{a}{The local baseline is the ASM observation closest
in time to the pre-flare observations, occurring at least one hour
earlier.} \tablenotetext{b}{Includes epoch 50590.16 for which some
X-ray observations exist within a range spanned by the 10\%
uncertainty that was assigned to all flare start time estimates in
Section 2.1}
\end{table*}

\par Figures 1 - 11 plot the (unabsorbed) intrinsic X-ray luminosity, $L_{\mathrm{intrinsic}}(1.2 - 50)$, light curves based on formulas
in equations (34) -(37) and a distance of 11 kpc. The intrinsic luminosity light
curves typically do not have the extreme variations of the raw Crab flux light
curves that appears in the same plots. This seems to indicate that
the intrinsic luminosity is considerably less variable than would be
assumed by just looking at raw ASM count rates, in general.
\subsection{X-ray Properties Before and During Flares} The formulas
in equations (34) - (37) are used to estimate the 1.2 keV to 50 keV
intrinsic X-ray luminosity, $L_{\mathrm{intrinsic}}(1.2 - 50)$, just
before and the time averaged luminosity during flares
($L_{\mathrm{pre-flare}}$, $L_{\mathrm{rise}}$, hereafter). The
results and the associated errors are compiled in Table 3. A
fiducial distance of 11 kpc is chosen for the table. Column (3) is
the date of the closest RXTE observation to the start of the major
ejection that is obtained from Figures 1 - 11 and is previously
tabulated in Table 1. Column (2) is the ASM observation with a date
that precedes the pre-flare X-ray date in column (3) that is closest
in time with a time difference of at least one hour. We call this a
(local) baseline from which the pre-flare flux varies
($L_{\mathrm{5-10 hr}}$, hereafter). It is close enough in time to
be plausibly causally connected to the pre-flare X-ray luminosity.
In column (1) we look at ASM observations to establish another more
distant baseline, the luminosity 24 hours ($L_{\mathrm{24 hr}}$,
hereafter) before the pre-flare RXTE observation of column (3). This
may or may not be causally connected. The luminosity in columns (4)
- columns (7) are computed from equations (34) - (37). ASM
observations during the actual rise of the flares exist for only 6
epochs. The ASM fluxes tend to be extremely time variable during the
rise of the flares, so a time average is computed in column (7). The
last 3 columns tabulate the hardest flux detected by ASM, bin 3.
Counts were converted to flux with equation (30c).
\par We list the conclusions that are drawn from the results in Table 3.
\begin{enumerate}
\item Columns (6) and (7) of Table 3 indicate that the time averaged X-ray
luminosity decreases during the plasmoid ejection, even though it is
highly variable with strong peaks that can exceed the pre - flare
levels.
\item Columns (6) and (7) of Table 3 indicate that $L_{\mathrm{rise}}$, is correlated with $L_{\mathrm{pre-flare}}$ (see Table 5).
\item $L_{\mathrm{pre-flare}}$ is significantly elevated relative to the archival distribution
of X-ray luminosity for GRS~1915+105. A Kolmogorov-Smirnov test
shows that $L_{\mathrm{pre-flare}}$ tends to be of larger values than would be expected if they
were representative of the distribution of luminosity that is
derived from the complete ASM  database of GRS~1915+105 observations
at the 98.0\% significance level. This is evident from the histogram
in Figure 16.
\item Columns (9) and (10) indicate a suppression of hard flux during
plasma ejection. This might indicate a softening of the spectrum, but the ASM data is not good enough to determine this.
The change in hard flux in of itself is insufficient to
explain the preponderance of the decreased total flux during the flare rise.
\end{enumerate}
\par Figure 16 is based on over 99.5\% of the archival ASM database that satisfy the softness criteria in
equation (32) for which the estimators in equations (34) and (35) are valid (over 76,000 epochs). In the bottom
frame of Figure 16, we explore the postulate of \citet{dha04} that a
dip in the X-ray luminosity precedes the flare launch when the time
scale defining the word "precede" is on the order of days. To this
end, we compared $L_{\mathrm{24 hr}}$ ($\sim$ 30 hours before flare ejection)
to the distribution of the ASM archival archive.
A Kolmogorov-Smirnov test shows that the intrinsic luminosity 30
hours before flare launch is depressed relative to the archival
distribution with a 96.9\% statistical significance. What is
particularly striking about this distribution is that every X-ray
state 30 hours before flare ejection lies within a very narrow range of
intrinsic X-ray luminosity, $\sim$ 4 - 5 $\times 10^{38}$ ergs/s.
\subsection{PCA Observations Near Flare Ejections}
The closest RXTE/PCA observations (before) the onset of a major
flare occurred on 51602.330--51602.361 for a flare start time
51602.506. GRS 1915+105 is in a class $\chi$ over the entire
observation. The 1.5-50 keV unabsorbed  flux is  $\gtrsim 4.2 \times
10^{-8}$~erg/cm$^2$/s indicating a rather bright occurrence for this
class. This is, however, typical of the luminous "high hard states"
or high plateau states mentioned in \citet{fuc03}. We also note a
slight decrease of the source unabsorbed flux over the last 2000~s
of the observation.
\par We, also, explored the temporal properties of the source,  verifying first through
inspection of a dynamical power density spectrum (PDS),  that they
were steady over the observation, as expected in a steady class
$\chi$ observation (see e.g. Rodriguez et al. 2008b). We, then,
extracted a PDS over the entire observation, and fitted it between
$0.01$ and $67$~Hz. Above typically $20$~Hz the PDS is consistent
with white noise and was modeled with a constant with a value of 2.
\par A good fit was achieved with 4 Lorentzian (on top of the white noise), three of then with broad profile
(one centered at 0 Hz), and a narrow feature indicative of the
presence of a low frequency quasi-periodic oscillation (QPO).
Inspection of the PDS, however, showed the presence of a slight
residual (a kind of QPO shoulder)  near the QPO. Adding another thin
Lorentzian improved the fit significantly. The total RMS amplitude
is 36.6~\%. The QPO has a frequency of $\sim 4.25$~Hz and an RMS
amplitude of $\sim6.7$~\%, while the additional feature has a
frequency $\sim 4.79$~Hz and an RMS amplitude of $\sim4.7$~\%.
Similar complicated QPO profiles have been seen in other sources
(e.g. XTE J1859+226, \citet{rod11} and references therein), and may
indicate the inadequacy of Lorentzian fitting in specific cases.
While the total RMS amplitude is rather usual for this type of
class, we note that the QPO amplitude is, on the contrary, rather
low, since class $\chi$ QPO  can have amplitude of 10--15\%
\citep{rod04}.
\par We also consider the RXTE/PCA observation
covering 51003.174--51003.2245 that occurs near the end of the
injection epsiode (possibly after) corresponding to the flare 51003.07. GRS 1915+105 is also in a
class $\chi$ over the whole observation. The 1.5--50 keV unabsorbed
flux is $3.1\times10^{-8}$~erg/cm$^2$/s. The total RMS amplitude is
$\sim 43.7$~\% somewhat higher than in the previous observation. The
PDS shows the presence of 2 QPOs harmonically related. The
fundamental has a frequency of $\sim 2.46$~Hz and an RMS amplitude
$\sim 12.1$~\%, while its harmonic is at $\sim 4.85$~Hz and has an
RMS amplitude $\sim 5.0$~\%.

\begin{figure}
\begin{center}
\includegraphics[width=90 mm, angle= 0]{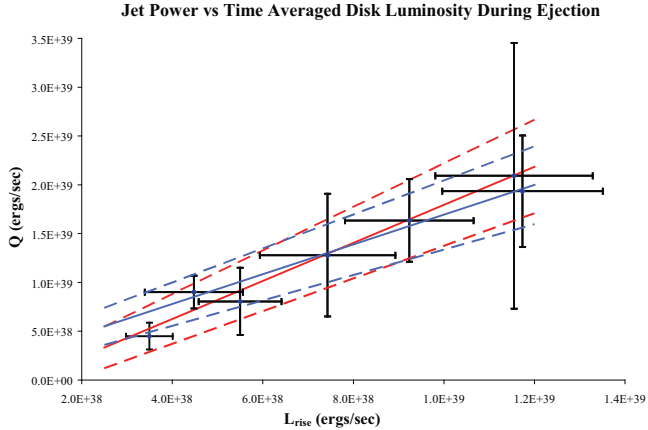}

\caption{\footnotesize{The correlation between the time averaged
X-ray luminosity during a major flare ejection (from Table 3) and
the estimated time averaged power required to eject the flare, $Q$,
that is taken from Table 2. The solid blue line is the weighted
least squares fit from Table 5. The dashed blue lines indicate the
standard error to the fit from Table 5. The point at the far left of
the plot represents epoch 50590.16. Formally X-ray observations
precede the flare start time, but some X-ray observations exist
within a range spanned by the 10\% uncertainty that was assigned to
all flare start time estimates in Section 2.1. The red lines
indicate the fit with errors from Table 5 if the flare 50590.16 is
included in the scatter plot, the blue lines are without flare
50590.16 included in the fit.}}
\end{center}
\end{figure}
\section{The Relationship Between X-ray
Luminosity and the Ejection Power} The reason for developing the
estimates of $Q$ and $L_{\mathrm{intrinsic}}(1.2 - 50)$ was to
provide a physical insight into the the provocative correlation
between pre-flare X-ray flux and peak optically thin flare 2.3 GHz
flux density in Figure 14. In Table 5, we list all the potentially
interesting correlations that we found. The first two columns X and
Y are the variables for a weighted least squares fit to $Y = mX + b$ with errors in
both variables using the methods of \citet{ree89}.
The estimated m and b are listed in columns 4 and 5. The next two
columns are the correlation coefficients and the probability that
the data scatter occurs by random chance.

\begin{table*}
\caption{Correlation Analysis} {\tiny\begin{tabular}{cccccccc}
\tableline\rule{0mm}{3mm}
X &  Y  & D  & m & b  & Pearson  & Spearman Rank & Figure \\
 &   &  & slope \tablenotemark{g}  & intercept \tablenotemark{g}   & Coefficient/ & Coefficient/ &   \\
  &   & (kpc) &  &  & (Probability)\tablenotemark{c}  & (Probability) \tablenotemark{c}   & \\
   \tableline \rule{0mm}{3mm}
Log(X-ray & Log[S] & &  & & & &  \\
Flux)\tablenotemark{d} & \tablenotemark{a}      & ... & $1.45 \pm 0.17$ & $2.27 \pm 0.04$  & 0.938/($1.90\times 10^{-5}$)   & 0.879/($5.20\times 10^{-3}$) & 14 \\
Log(X-ray & Log[S]& &  & & & &  \\
Flux)\tablenotemark{d} & \tablenotemark{f}      & ... & $1.40 \pm 0.11$  & $2.31 \pm 0.03$  &  0.966/($1.28\times 10^{-6}$)   & 0.952/($2.60\times 10^{-3}$)  & ...\\
$L_{\mathrm{pre-flare}}$\tablenotemark{b} & $E_{\mathrm{plasmoid}}$ \tablenotemark{e} & 11 & $7.37\times 10^{3} \pm  8.74 \times 10^{2} $ & $-2.82\times 10^{42} \pm 5.27\times 10^{41}$ & 0.892/($2.20\times 10^{-4}$)   & 0.954/($2.60\times 10^{-3}$) & ...\\
$L_{\mathrm{pre-flare}}$\tablenotemark{b} & $E_{\mathrm{plasmoid}}$ \tablenotemark{e} & 10.5 & $1.71\times 10^{3} \pm  2.00 \times 10^{2} $ & $-5.91\times 10^{41} \pm 1.10\times 10^{41}$ & 0.912/($9.32\times 10^{-5}$)   & 0.927/($3.20\times 10^{-3}$) & ...\\
$L_{\mathrm{pre-flare}}$\tablenotemark{b} & $E_{\mathrm{plasmoid}}$ \tablenotemark{e} & 10 & $8.75\times 10^{2} \pm  9.90 \times 10^{1} $ & $-2.75\times 10^{41} \pm 4.96\times 10^{40}$ & 0.912/($9.32\times 10^{-5}$)   & 0.927/($3.20\times 10^{-3}$) & ...\\
$L_{\mathrm{pre-flare}}$\tablenotemark{b} & $Q$ \tablenotemark{b} & 11 & $1.70 \pm 0.21$ & $-3.49 \times 10^{38} \pm 1.31 \times 10^{38}$  & 0.955/($4.81\times 10^{-6}$)   & 0.973/($2.00\times 10^{-3}$) & 17 \\
$L_{\mathrm{pre-flare}}$\tablenotemark{b} & $Q$ \tablenotemark{b} & 10.5 & $0.24 \pm 0.04$ & $-4.12 \times 10^{37} \pm 2.24 \times 10^{37}$  & 0.880/($3.51\times 10^{-4}$)   & 0.936/($3.00\times 10^{-3}$) & ... \\
$L_{\mathrm{pre-flare}}$\tablenotemark{b} & $Q$ \tablenotemark{b} & 10 & $0.10 \pm 0.01$ & $-1.52 \times 10^{37} \pm 0.73 \times 10^{37}$  & 0.906/($1.22\times 10^{-4}$)   & 0.955/($2.60\times 10^{-3}$) & ... \\
$L_{\mathrm{pre-flare}}$\tablenotemark{b} & $Q+ L_{\mathrm{rise}}$ \tablenotemark{b} & 11 & $3.00 \pm 0.33$ & $-8.60 \times 10^{38} \pm 2.99 \times 10^{38}$  & 0.955/($3.06\times 10^{-3}$)   & 0.886/($4.66\times 10^{-2}$) & ... \\
$L_{\mathrm{pre-flare}}$\tablenotemark{b} & $Q+ L_{\mathrm{rise}}$ \tablenotemark{b} & 10.5 & $1.41 \pm 0.23$ & $-3.78 \times 10^{38} \pm 1.96 \times 10^{38}$  & 0.938/($5.62\times 10^{-3}$)   & 0.886/($4.66\times 10^{-2}$) & ... \\
$L_{\mathrm{pre-flare}}$\tablenotemark{b} & $Q+ L_{\mathrm{rise}}$ \tablenotemark{b} & 10 & $1.25 \pm 0.19$ & $-3.06 \times 10^{38} \pm 1.50 \times 10^{38}$  & 0.955/($3.10\times 10^{-3}$)   & 0.886/($4.66\times 10^{-2}$) & ... \\
$L_{\mathrm{pre-flare}}$\tablenotemark{b,j} & $Q+ L_{\mathrm{rise}}$ \tablenotemark{b,j} & 11 & $2.35 \pm 0.24$ & $-2.15 \times 10^{38} \pm 1.59 \times 10^{38}$  & 0.952/($9.61\times 10^{-4}$)   & 0.893/($2.86\times 10^{-2}$) & ... \\
$L_{\mathrm{pre-flare}}$\tablenotemark{b,j} & $Q+ L_{\mathrm{rise}}$ \tablenotemark{b,j} & 10.5 & $0.99 \pm 0.15$ & $1.09 \times 10^{37} \pm 9.49 \times 10^{37}$  & 0.937/($1.82\times 10^{-3}$)   & 0.929/($2.26\times 10^{-2}$) & ... \\
$L_{\mathrm{pre-flare}}$\tablenotemark{b,j} & $Q+ L_{\mathrm{rise}}$ \tablenotemark{b,j} & 10 & $0.87 \pm 0.13$ & $1.63 \times 10^{37} \pm 7.85 \times 10^{37}$  & 0.944/($1.38\times 10^{-3}$)   & 0.929/($2.26\times 10^{-2}$) & ... \\
$L_{\mathrm{rise}}$\tablenotemark{b} & $E_{\mathrm{plasmoid}}$ \tablenotemark{e} & 11 & $8.09\times 10^{3} \pm  2.49 \times 10^{3} $ & $-1.57\times 10^{42} \pm 1.49\times 10^{42}$ & 0.775/($6.99\times 10^{-2}$)   & 0.886/($4.66\times 10^{-2}$) & ...\\
$L_{\mathrm{rise}}$\tablenotemark{b} & $E_{\mathrm{plasmoid}}$ \tablenotemark{e} & 10.5 & $1.72\times 10^{3} \pm  4.51 \times 10^{2} $ & $-2.82\times 10^{41} \pm 2.46\times 10^{41}$ & 0.820/($4.59\times 10^{-2}$)   & 0.886/($4.66\times 10^{-2}$) & ...\\
$L_{\mathrm{rise}}$\tablenotemark{b} & $E_{\mathrm{plasmoid}}$ \tablenotemark{e} & 10 & $9.11\times 10^{2} \pm  2.17 \times 10^{2} $ & $-1.30\times 10^{41} \pm 1.17\times 10^{41}$ & 0.872/($1.18\times 10^{-2}$)   & 0.886/($4.66\times 10^{-2}$) & ...\\
$L_{\mathrm{rise}}$\tablenotemark{b,j} & $E_{\mathrm{plasmoid}}$ \tablenotemark{e,j} & 11 & $1.04\times 10^{4} \pm  2.05 \times 10^{3} $ & $-3.14\times 10^{42} \pm 1.00\times 10^{42}$ & 0.832/($2.01\times 10^{-2}$)   & 0.929/($2.26\times 10^{-2}$) & ...\\
$L_{\mathrm{rise}}$\tablenotemark{b,j} & $E_{\mathrm{plasmoid}}$ \tablenotemark{e,j} & 10.5 & $2.40\times 10^{3} \pm  4.80 \times 10^{2} $ & $-7.13\times 10^{41} \pm 2.26\times 10^{41}$ & 0.869/($1.11\times 10^{-2}$)   & 0.929/($2.26\times 10^{-2}$) & ...\\
$L_{\mathrm{rise}}$\tablenotemark{b,j} & $E_{\mathrm{plasmoid}}$ \tablenotemark{e,j} & 10 & $9.12\times 10^{2} \pm  2.54 \times 10^{2} $ & $-2.49\times 10^{41} \pm 8.70\times 10^{40}$ & 0.907/($4.76\times 10^{-3}$)   & 0.929/($2.26\times 10^{-2}$) & ...\\
$L_{\mathrm{rise}}$\tablenotemark{b} & $Q$ \tablenotemark{b} & 11 & $1.52 \pm 0.22$ & $1.68 \times 10^{38} \pm 1.36 \times 10^{38}$  & 0.980/($5.95\times 10^{-4}$)   &  0.886/($4.66\times 10^{-2}$) & 18 \\
$L_{\mathrm{rise}}$\tablenotemark{b} & $Q$ \tablenotemark{b} & 10.5 & $0.20 \pm 0.03$ & $2.62 \times 10^{37} \pm 1.88 \times 10^{37}$  & 0.931/($6.89\times 10^{-3}$)   &  0.772/($8.36\times 10^{-2}$) & ... \\
$L_{\mathrm{rise}}$\tablenotemark{b} & $Q$ \tablenotemark{b} & 10 & $0.08 \pm 0.01$ & $8.50 \times 10^{36} \pm 6.05 \times 10^{36}$  & 0.949/($3.77\times 10^{-3}$)   &  0.886/($4.66\times 10^{-2}$) & ... \\
$L_{\mathrm{rise}}$\tablenotemark{b,j} & $Q$ \tablenotemark{b,j} & 11 & $1.95 \pm 0.28$ & $-1.55 \times 10^{38} \pm 1.43 \times 10^{38}$  & 0.984/($6.23\times 10^{-5}$)   &  0.929/($2.26\times 10^{-2}$) & 18 \\
$L_{\mathrm{rise}}$\tablenotemark{b,j} & $Q$ \tablenotemark{b,j} & 10.5 & $0.27 \pm 0.05$ & -$2.04 \times 10^{37} \pm 2.24 \times 10^{37}$  & 0.948/($1.13\times 10^{-3}$)   &  0.857/($3.58\times 10^{-2}$) & ... \\
$L_{\mathrm{rise}}$\tablenotemark{b,j} & $Q$ \tablenotemark{b,j} & 10 & $0.11 \pm 0.01$ & $-6.53 \times 10^{36} \pm 7.68 \times 10^{36}$  & 0.961/($5.76\times 10^{-4}$)   &  0.929/($2.26\times 10^{-2}$) & ... \\
$\Delta L_{6hr}$ \tablenotemark{h} & $Q$ \tablenotemark{b} & 11 & $2.22 \pm 0.33$ & $5.95 \times 10^{38} \pm 8.32 \times 10^{37}$  & 0.937/($6.33\times 10^{-5}$)   &  0.915/($6.00\times 10^{-3}$) & ... \\
$\Delta L_{6hr}$ \tablenotemark{h} & $Q$ \tablenotemark{b}& 10.5 & $0.31 \pm 0.07$ & $8.25 \times 10^{37} \pm 1.54 \times 10^{37}$  & 0.864/($1.28\times 10^{-3}$)   &  0.891/($7.40\times 10^{-3}$) & ... \\
$\Delta L_{6hr}$ \tablenotemark{h} & $Q$ \tablenotemark{b}& 10 & $0.12 \pm 0.03 $ & $2.93 \times 10^{37} \pm 5.09 \times 10^{36}$  & 0.893/($5.02\times 10^{-4}$)   &  0.927/($5.20\times 10^{-3}$) & ... \\
$\Delta L_{24hr}$ \tablenotemark{i} & $Q$ \tablenotemark{b} & 11 & $1.79 \pm 0.16$ & $4.05 \times 10^{38} \pm 4.29 \times 10^{37}$  & 0.973/($5.32\times 10^{-7}$)   &  0.964/($2.20\times 10^{-3}$) & 19 \\
$\Delta L_{24hr}$ \tablenotemark{i} & $Q$ \tablenotemark{b}& 10.5 & $0.25 \pm 0.04$ & $5.48 \times 10^{37} \pm 9.49 \times 10^{36}$  & 0.893/($2.16\times 10^{-4}$)   &  0.918/($3.60\times 10^{-3}$) & ... \\
$\Delta L_{24hr}$ \tablenotemark{i} & $Q$ \tablenotemark{b}& 10 & $0.12 \pm 0.01 $ & $1.68 \times 10^{37} \pm 3.66 \times 10^{36}$  & 0.921/($5.65\times 10^{-5}$)   &  0.936/($3.00\times 10^{-3}$) & ... \\
$L_{\mathrm{pre-flare}}$\tablenotemark{b} & $L_{\mathrm{rise}}$ \tablenotemark{b} & \tablenotemark{k} & $1.16 \pm 0.19$ & $-3.57 \times 10^{38} \pm 1.72 \times 10^{37}$  & 0.955/($2.73\times 10^{-3}$)   &  0.943/($3.48\times 10^{-2}$) & ... \\
$L_{\mathrm{pre-flare}}$\tablenotemark{b,j} & $L_{\mathrm{rise}}$ \tablenotemark{b,j} & \tablenotemark{k} & $0.84 \pm 0.12$ & $-3.79 \times 10^{37} \pm 9.01 \times 10^{37}$  & 0.953/($8.93\times 10^{-4}$)   &  0.964/($1.78\times 10^{-2}$) & ... \\
\end{tabular}}
\tablenotetext{a}{\tiny{Peak observed $S_{\nu=2.3\mathrm{GHz}}$ in
mJy}} \tablenotetext{b}{\tiny{cgs units}}
\tablenotetext{c}{\tiny{Probability of occurring by random chance}}
\tablenotetext{d}{\tiny{Crab units}} \tablenotetext{e}{\tiny{The
energy of the ejected plasmoid in ergs}}
\tablenotetext{f}{\tiny{Estimated $S_{\nu=2.3\mathrm{GHz}}(\tau
=0.1)$ in mJy}} \tablenotetext{g}{\tiny{Weighted least squares fit
with errors in both variables to $Y = mX +b$ computed per
\citet{ree89}}} \tablenotetext{h}{\tiny{$L_{\mathrm{pre-flare}} -
L_{\mathrm{5-10 hr}}$ in erg/s}}
\tablenotetext{i}{\tiny{$L_{\mathrm{pre-flare}} - L_{\mathrm{24
hr}}$ in erg/s}} \tablenotetext{j}{\tiny{Includes epoch 50590.16 for
which some X-ray observations exist within a range spanned by the
10\% uncertainty that was assigned to all flare start time estimates
in Section 2.1}} \tablenotetext{k}{\tiny{Independent of the distance
to source.}}
\end{table*}
\par The analog of Figure 14 in terms of physical estimates is $Q$ as a function of
$L_{\mathrm{pre-flare}}$ that is plotted in Figure 17 with the data from
Tables 2 and 3 and the curve fit and errors from Table 5.
Notice from Table 5 that the correlation is much improved from Figure 14.
The most direct indicator of the flare launch physics is $L_{\mathrm{rise}}$.
Unfortunately, there are only 6
major flares with X-ray data during launch. In spite of the small
number statistics, the chance to probe this most fundamental
indicator of the physics of major flare ejection makes it a
worthwhile exercise to look for trends in the data. The X-ray light
curves during flares is often highly time variable in Figures 1 - 11,
similar to the post ejection X-ray state discussed in \citet{tru07}.
By contrast, the estimators of flare energy and power in Section 3
are only time averages. Thus, we must use time averaged X-ray luminosity during the ejection,
$L_{\mathrm{rise}}$, in order to make relevant comparisons. This is affected by the aforementioned
variability and the sparse and nonuniform time sampling during the
brief flare rise. Thus, depending on the time sampling there can be
significant uncertainty in this average that is best dealt with by
larger samples. The scatter plot with the fits from Table 5 are plotted in Figure 18.
Noting the strong variability in the X-ray luminosity in Figures 1 - 11 during
flare launching, it seems likely that the instantaneous power
required to launch the plasmoid is highly variable as well. This
seems mirrored in the VLBA image in Figure 13 which is consistent
with a strong abrupt burst (or bursts) followed by weaker ejecta. It
is not known if these would appear in or out of phase with the X-ray
bursts without finer time resolution in the radio data sampling.
\begin{figure}
\begin{center}
\includegraphics[width=90 mm, angle= 0]{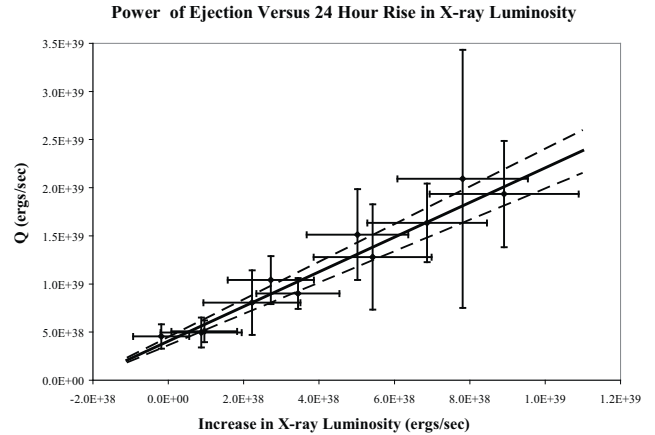}
\caption{\footnotesize{A plot of the correlation that exists between
the increase in intrinsic X-ray luminosity one day preceding the
ejection (from Table 3) and the estimated time averaged power of the
ejection, $Q$. The plot is for the special case D = 11 kpc. The
solid black line is the weighted least squares fit from Table 5. The
dashed black lines indicate the standard error to the fit from Table
5.}}
\end{center}
\end{figure}
\par In Figure 19, we plot the increase in the
$L_{\mathrm{intrinsic}}(1.2 - 50)$ in the last day ($ L_{\mathrm{pre-flare}}- L_{\mathrm{24 hr}}$) preceding
the plasma ejection (from Table 3) versus $Q$. There is a significant
correlation that extends continuously to the weaker flares in which
there is actually a decrease in luminosity before the ejection
instead of an increase. The correlation in Figure 19 is redundant with that in Figure 17. It arises as a consequence of the fact that the
X-ray luminosity one day before flare launch is confined to within a
tight window as noted in regards to Figure 16, $\sim$ 4 - 5 $\times
10^{38}$ ergs/s (see Table 3).
\par In summary, Tables 3 and 5 indicate the following evolution of the X-ray luminosity before and during flare launch.
For the flares in our study, one day before the ejection, the system
is in a low luminosity state with $L_{\mathrm{24 hr}} \sim  4 -
5\times 10^{38}$ ergs/s. We find that major flares can be launched
if the X-ray luminosity decreases or increases over the next 24
hours. The power and energy of the ejection scales with
$L_{\mathrm{pre-flare}}$. Our lone pointed observation, 3 to 4 hours
before a major flare (epoch 51602.506), shows no spectral signature
that indicates that an ejection is imminent. Table 3 and Figure 19
indicate that for strong flares $L_{\mathrm{intrinsic}}(1.2 - 50)$
continues to increase from roughly a day before launch into the
final hours preceding the ejection. From Table 5 when the flare is
launched, the time averaged power, $Q$ + $L_{\mathrm{rise}}$, is
highly correlated with $L_{\mathrm{pre-flare}}$, with near equality
if $D \approx 10.5 $ kpc.

\section{Discussion}We have presented the most
comprehensive possible study of the X-ray properties of GRS~1915+105
just prior to and during major flare ejection given the state of the
available data. Our primary empirical result that is model
independent is the strong correlation between the 1.2--12 keV
X-ray flux 0 to 4 hours before a major flare ejection and the peak
optically thin 2.3 GHz flux density of the flare. Most of the
article was spent developing this result into one involving basic
physical entities. In particular, we deduce the following results
\begin{enumerate}
\item The time averaged power required to launch the major flares, $Q$, is highly correlated with
$L_{\mathrm{pre-flare}}$ (Table 5).
\item $L_{\mathrm{pre-flare}}$ is significantly elevated relative to the archival distribution
of $L_{\mathrm{intrinsic}}(1.2 - 50)$ for GRS~1915+105. (Figure 16, Section 4).
\item The time averaged power required to launch the major flares, $Q$, is correlated with $L_{\mathrm{rise}}$
(Table 5).
\item From Table 5, $Q$ + $L_{\mathrm{rise}}$ is
highly correlated with $L_{\mathrm{pre-flare}}$, with near equality
if $D \approx 10.5 $ kpc.
\item Columns (4) - (6) of Table 3 indicate that $L_{\mathrm{intrinsic}}(1.2 - 50)$
increases in the hours that precede the stronger flares in the
sample. There is a correlation between the amount of
$L_{\mathrm{intrinsic}}(1.2 - 50)$ increase (that extends to the
decrease regime) and the strength of the subsequent flare (see
Figure 19 and Section 5).
\end{enumerate}
These findings are consistent with the interpretation that accretion rate
is directly proportional to flare power, but elevated accretion in
of itself does not guarantee that a major flare is ejected (in fact
they usually are not).

\par Our data set spans nearly the full range of major flare power that
covers a factor $\sim 30$ in estimated flare energy and a
factor of $\sim 5$ in time averaged flare ejection power, $Q$.
Based on the 2.3 GHz flux density, these flares approach energies and
powers that rival the most powerful known flares since 1995.

\par Further radio and X-ray
monitoring should be performed with high time resolution. This will allow the
correlations described in this manuscript to be tested with more
rigor and deeper physical connections should be revealed. From a
physical point of view, the most critical to understand would be the
connection between the variable X-ray luminosity during the launch
phase and the identification of individual flux density peaks in
high frequency VLBI images.
\begin{acknowledgements}
This article benefitted from a very knowledgable referee who made many important improvements and suggestions.
We would like to thank Vivek Dhawan for generously sharing his deep
understanding of radio observations and his important data. We also
are grateful for the useful comments on the manuscript generously provided
by S. Markoff and S. Corbel. JR acknowledges partial funding from the European FP7 grant agreement
number ITN 215212 ``Black Hole Universe", and the hospitality of ESO
(Garching, Germany) where part of this work was done.
\end{acknowledgements}

\end{document}